\documentclass[preprint]{imsart}

\RequirePackage[OT1]{fontenc}
\RequirePackage{amsthm,amsmath}
\RequirePackage{natbib}
\RequirePackage[colorlinks,citecolor=blue,urlcolor=blue]{hyperref}

\startlocaldefs
\numberwithin{equation}{section}
\theoremstyle{plain}

\endlocaldefs

\usepackage{booktabs}
\usepackage{soul}
\usepackage{amsmath,amsfonts,amssymb,amsbsy}
\usepackage{url}
\usepackage{xcolor}
\usepackage{footnote}
\usepackage{graphicx} %
\usepackage{caption} 
\usepackage{subcaption}
\usepackage{placeins} %
\usepackage[titletoc]{appendix} %
\usepackage{wrapfig}
\usepackage{listings}
\usepackage{verbatim}

\def\codesdir{codes/} %

\usepackage{ifthen}
\usepackage{tikz,pgfplots}
\usetikzlibrary{matrix}
\usetikzlibrary{calc}
\newlength{\figurewidth}
\newlength{\figureheight}

\def\figpdfdir{fig/} %
\def\figtikzdir{tikz/} %

\newcommand{\minput}[2][]{
\ifthenelse{\equal{#1}{pdf}}
	{ \includegraphics{\figpdfdir #2} }
	{ \tikzset{external/remake next} \tikzsetnextfilename{#2} \input{\figtikzdir #2} }
}

\usetikzlibrary{external}
\tikzset{external/system call={lualatex
	\tikzexternalcheckshellescape -halt-on-error -interaction=batchmode
	-jobname "\image" "\texsource"}}
\newcommand{\vc}[1] { \mathbf{#1} } %
\newcommand{\vs}[1] { \boldsymbol{#1} }
\newcommand{\tp}{\mathsf{T}}
\newcommand{\ti}[1] { \tilde{#1} } 
\newcommand{\mc}[1] { \mathcal{#1} } 
\newcommand{\tx}[1] { \text{#1} } 
\newcommand{\given} { \,|\, }

\newcommand{\mean}[2][] { \mathrm{E}_{#1} {\left(#2\right)} }
\newcommand{\var}[1] { \mathrm{Var} {\left(#1\right)} }

\newcommand{\diag}[1] { \mathrm{diag} {\left(#1\right)} }

\newcommand{\doo} {\partial}

\newcommand{\Normal}[1] { \mathrm{N} {\left(#1\right)}  }
\newcommand{\halfNormal}[1] { \mathrm{N}^+ {\left(#1\right)}  } 
\newcommand{\Student}[2][] { t_{#1} {\left(#2\right)} }
\newcommand{\halfStudent}[2][] { t_{#1}^+ {\left(#2\right)} }

\newcommand{\InvGamma}[1] { \mathrm{Inv\text{-}Gamma} {\left(#1\right)} }

\newcommand{\halfCauchy}[1] { \mathrm{C}^+ {\left(#1\right)} }
\newcommand{\Beta}[1] {\mathrm{Beta}{\left(#1\right)} }
\newcommand{\Ber}[1] {\mathrm{Ber}{\left(#1\right)} }

\AtBeginDocument{%
   \def\MR#1{}
}

\begin{document}

\begin{frontmatter}
\title{Sparsity information and regularization in the horseshoe and other \\shrinkage priors}%
\runtitle{ Sparsity and regularization in the horseshoe prior}

\begin{aug}
\author{\fnms{Juho} \snm{Piironen}\ead[label=e1]{juho.piironen@aalto.fi}}
\and
\author{\fnms{Aki} \snm{Vehtari}\ead[label=e2]{aki.vehtari@aalto.fi}}

\address{
Helsinki Institute for Information Technology, HIIT \\ Department of Computer Science, Aalto University\\
\printead{e1,e2}
}

\runauthor{J. Piironen and A. Vehtari }

\affiliation{Aalto University}

\end{aug}

\begin{abstract}
The horseshoe prior has proven to be a noteworthy alternative for sparse Bayesian estimation, but has previously suffered from two problems.
First, there has been no systematic way of specifying a prior for the global shrinkage hyperparameter based on the prior information about the degree of sparsity in the parameter vector.
Second, the horseshoe prior has the undesired property that there is no possibility of specifying separately information about sparsity and the amount of regularization for the largest coefficients, which can be problematic with weakly identified parameters, such as the logistic regression coefficients in the case of data separation.
This paper proposes solutions to both of these problems.
We introduce a concept of effective number of nonzero parameters, show an intuitive way of formulating the prior for the global hyperparameter based on the sparsity assumptions, and argue that the previous default choices are dubious based on their tendency to favor solutions with more unshrunk parameters than we typically expect a priori.
Moreover, we introduce a generalization to the horseshoe prior, called the regularized horseshoe, that allows us to specify a minimum level of regularization to the largest values.
We show that the new prior can be considered as the continuous counterpart of the spike-and-slab prior with a finite slab width, whereas the original horseshoe resembles the spike-and-slab with an infinitely wide slab.
Numerical experiments on synthetic and real world data illustrate the benefit of both of these theoretical advances.
\end{abstract}

\begin{keyword}[class=MSC]
\kwd[Primary ]{62F15}
\end{keyword}

\begin{keyword}
\kwd{Bayesian inference}
\kwd{sparse estimation}
\kwd{shrinkage priors}
\kwd{horseshoe prior}
\end{keyword}
\end{frontmatter}

\section{Introduction}

This paper deals with sparse Bayesian estimation and is an extension to our earlier work \citep{piironen2017b}.
We consider statistical models with a large number of parameters $\vs \theta=(\theta_1,\dots,\theta_D)$  but so that it is reasonable to assume that only some of them are far from zero.
A typical example -- and also the case we will mostly focus in this paper -- is a regression or classification problem with a large number of predictor variables out of which we expect only a few to be relevant and therefore have a regression coefficient distinguishable from zero.

A vast number of different estimators, both Bayesian and non-Bayesian, have been proposed for these problems.
In the non-Bayesian literature the sparse problems are typically handled by Lasso \citep{tibshirani1996} or one of its generalizations~\citep[for an overview, see e.g.,][]{hastie2015book}.
We focus on the probabilistic approach and carry out full Bayesian inference on the problem.

Two prior choices dominate the Bayesian literature: two component discrete mixture priors known as the spike-and-slab \citep{mitchell1988,george1993}, and a variety of continuous shrinkage priors \citep[see e.g.,][and references therein]{polson2011}.
The spike-and-slab prior is intuitively appealing as when the spike is taken to be a delta-spike in the origin, it is equivalent to Bayesian model averaging (BMA)~\citep{hoeting1999} over the variable combinations, and often has good performance in practice.
The disadvantages are that the results can be sensitive to prior choices (slab width and prior inclusion probability) and that the posterior inference can be computationally demanding with a large number of variables, due to the huge model space.
The inference could be sped up by analytical approximations using either expectation propagation (EP) \citep{hernandezlobato2010,hernandezlobato2015} or variational inference (VI) \citep{titsias2011}, but this comes at the cost of a substantial increase in the amount of analytical work needed to derive the equations separately for each model and a more complex implementation.

The continuous shrinkage priors on the other hand are easy to implement, provide convenient computation using generic sampling tools such as Stan \citep{stan_manual}, and can yield as good or better results.
A particularly interesting example is the horseshoe prior \citep{carvalho2009,carvalho2010}
\begin{align}
\begin{split}
	\theta_j \given \lambda_j,\tau &\sim \Normal{0, \tau^2 \lambda_j^2}\,, \\
	\lambda_j &\sim \halfCauchy{0,1}\,, \quad j=1,\dots,D,
\end{split}
\label{eq:hs_generic}
\end{align}
which has shown comparable performance to the spike-and-slab prior in a variety of examples where a sparsifying prior on the model parameters $\theta_j$ is desirable \citep{carvalho2009,carvalho2010,polson2011}.
The horseshoe is one of the so called global-local shrinkage priors, meaning that there is a global hyperparameter $\tau$ that shrinks all the parameters towards zero, while the heavy-tailed half-Cauchy priors for the local hyperparameters $\lambda_j$ allow some $\theta_j$ to escape the shrinkage (see Sec.~\ref{sec:horseshoe} for more thorough discussion).

Despite its good performance in many problems, the horseshoe prior has previously suffered from two shortcomings. 
First, there has been no consensus on how to carry out inference for the global hyperparameter $\tau$ which determines the overall sparsity in the parameter vector $\vs \theta$ and therefore has a large impact on the results.
We prefer full Bayesian inference (see Sec.~\ref{sec:full_bayes_vs_point_estimation}) but the existing methodology has been lacking a systematic way of placing a prior for $\tau$ based on the information about the sparsity.
Second, the horseshoe prior has the undesired property that the parameters far from zero are not regularized at all (see Sec.~\ref{sec:regularized_horseshoe}).
While this is often considered as a key strength of the prior, it can be harmful especially when the parameters are only weakly identified by the data, for instance in the case of a flat likelihood due to separable data in logistic regression.

We propose a solution to both of these problems.
We introduce a concept of effective number of nonzero parameters $m_\tx{eff}$ (Sec.~\ref{sec:meff_and_tau}), derive analytically its relationship between the global shrinkage parameter $\tau$, and show an easy and intuitive way of formulating the prior for $\tau$ based on the prior information about the sparsity of $\vs \theta$.
Based on these theoretical considerations, we argue that the previously proposed default priors are dubious based on the prior they impose on $m_\tx{eff}$, and that they yield good results only when $\tau$ (and therefore $m_\tx{eff}$) is strongly identified by the data.
Moreover, we introduce a generalization of the horseshoe prior, called the regularized horseshoe, that operates otherwise similarly as the original horseshoe but allows specifying the regularization to the coefficients that are far from zero (see Sec.~\ref{sec:regularized_horseshoe}).
We show that the regularized horseshoe can be considered as the continuous counterpart of the spike-and-slab prior with a finite slab width, whereas the original horseshoe resembles the spike-and-slab with an infinitely wide slab.
The benefit of both of these theoretical advances will be illustrated with examples on synthetic and real world data (Sec.~\ref{sec:experiments}).

As a final remark, although we focus on the horseshoe in our discussion, we want to emphasize that both of these ideas could also be applied to other shrinkage priors, and several promising alternatives to the horseshoe have been proposed during the recent years \citep{bhattacharya2015, zhang2016, ghosh2017}.

\section{Horseshoe prior and its extension}
\label{sec:horseshoe_and_others}

This section discusses the horseshoe and its connection to the spike-and-slab prior (Section~\ref{sec:spike-and-slab}).
We also present an extension (Section~\ref{sec:regularized_horseshoe}) that both helps understanding the theoretical properties of the original horseshoe and -- as will be demonstrated in Section~\ref{sec:experiments} -- robustifies the prior and improves its practical performance.

\subsection{Horseshoe prior for linear regression}
\label{sec:horseshoe}

Consider the single output linear Gaussian regression model with several input variables, given by 
\begin{align}
\begin{split}
	y_i &= \vs \beta^\tp \vc x_i + \varepsilon_i, \quad \varepsilon_i \sim \Normal{0,\sigma^2}, \quad i=1,\dots,n \,, 
\end{split}
\label{eq:lgm}
\end{align}
where $\vc x$ is the $D$-dimensional vector of inputs, $\vs \beta$ contains the corresponding weights and $\sigma^2$ is the noise variance.
The horseshoe prior is set for the regression coefficients $\vs \beta = (\beta_1,\dots,\beta_D)$ 
\begin{align}
\begin{split}
	\beta_j \given \lambda_j,\tau &\sim \Normal{0, \tau^2 \lambda_j^2}, \\
	\lambda_j &\sim \halfCauchy{0,1} \,, \quad j=1,\dots,D.
\end{split}
\label{eq:hs}
\end{align}
If an intercept term $\beta_0$ is included in model~\eqref{eq:lgm}, we give it a relatively flat prior, because there is usually no reason to shrink it towards zero.
As discussed in the introduction, the horseshoe prior has been shown to possess several desirable theoretical properties and good performance in practice \citep{carvalho2009,carvalho2010,polson2011,datta2013,vanDerPas2014}.
The intuition is the following: the global parameter $\tau$ pulls all the weights globally towards zero, while the thick half-Cauchy tails for the local scales $\lambda_j$ allow some of the weights to escape the shrinkage.
Different levels of sparsity can be accommodated by changing the value of $\tau$: with large $\tau$ all the variables have very diffuse priors with very little shrinkage, but letting $\tau \rightarrow 0$ will shrink all the weights $\beta_j$ to zero.

The above can be formulated more formally as follows.
Let $\vc X$ denote the $n$-by-$D$ matrix of observed inputs and $\vc y$ the observed targets.
The conditional posterior for the coefficients $\vs \beta$ given the hyperparameters and data $\mc D = (\vc X, \vc y)$ can be written as
\begin{align*}
	p(\vs \beta \given \vs \Lambda, \tau,\sigma^2,\mc D) &= \Normal{\vs \beta \given \vs {\bar \beta}, \vs \Sigma},\\
	\vs {\bar \beta} &= \tau^2 \vs \Lambda \left( \tau^2 \vs \Lambda + \sigma^2 (\vc X^\tp \vc X)^{-1} \right)^{-1} \vs {\hat \beta}, \\
	\vs \Sigma &= (\tau^{-2}\vs \Lambda^{-1} + \frac{1}{\sigma^2} \vc X^\tp \vc X)^{-1},
\end{align*}
where $\vs \Lambda = \diag{\lambda_1^2,\dots,\lambda_D^2}$ and $\vs {\hat \beta} = (\vc X^\tp \vc X)^{-1} \vc X^\tp \vc y$ is the maximum likelihood solution (assuming the inverse exists).
If the predictors are uncorrelated with zero mean and variances $\var{x_j}=s_j^2$, then $\vc X^\tp \vc X \approx  n \, \diag{s_1^2,\dots,s_D^2}$, and we can approximate
\begin{align}
	\bar \beta_j = (1-\kappa_j)\hat \beta_j,
\label{eq:beta_mean}
\end{align}
where
\begin{align}
	\kappa_j = \frac{1}{1+n \sigma^{-2} \tau^2  s_j^2 \lambda_j^2}
\label{eq:kappa}
\end{align}
is the {\it shrinkage factor} for coefficient $\beta_j$.
The shrinkage factor describes how much coefficient $\beta_j$ is shrunk towards zero from the maximum likelihood solution ($\kappa_j=1$ meaning complete shrinkage and $\kappa_j=0$ no shrinkage).
From~\eqref{eq:beta_mean} and~\eqref{eq:kappa} it is easy to verify that $\vs {\bar \beta} \rightarrow 0$ as $\tau \rightarrow 0$, and $\vs {\bar \beta} \rightarrow \vs {\hat \beta}$ as $\tau \rightarrow \infty$.

\begin{figure}[t]
\centering
	\setlength{\figureheight}{0.15\textwidth} 
	\setlength{\figurewidth}{0.25\textwidth}
	\pgfplotsset{
	compat=newest,
	major tick length={0.03cm},
	x tick label style={font=\tiny},
	y tick label style={font=\tiny},
	ylabel style={text width=5em, rotate=-90, align=left},
	legend style={font=\tiny},
	} 
	\minput[pdf]{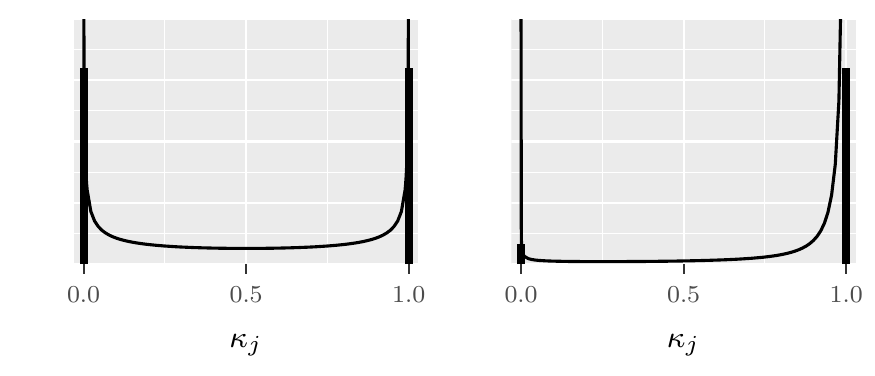}
	\caption{The continuous curves show the densities for the shrinkage factor~\eqref{eq:kappa} for the horseshoe prior~\eqref{eq:hs} when $a_j = \tau \sigma^{-1} \sqrt{n}\, s_j = 1$ (left) and when $a_j = 0.1$ (right). The bars denote the corresponding point mass function for the spike-and-slab  prior~\eqref{eq:spike-and-slab2} with infinite slab width $c\rightarrow \infty$, when $\pi/(1-\pi)=1$ (left) and $\pi/(1-\pi)=0.1$ (right). To aid visualization, the bars illustrating the point masses are scaled and show only the relative probability masses.}
	\label{fig:kappa_prior_hs}
\end{figure}

The result~\eqref{eq:kappa} holds for any prior that can be written as a scale mixture of Gaussians like~\eqref{eq:hs}, regardless of the prior for $\lambda_j$.
The horseshoe employs independent half-Cauchy priors for all $\lambda_j$, and for this choice one can show that, for fixed $\tau$ and $\sigma$, the shrinkage factor~\eqref{eq:kappa} follows the prior
\begin{align}
	p(\kappa_j\given \tau,\sigma) 
	&= \frac{1}{\pi} \frac{a_j}{(a_j^2-1)\kappa_j + 1}
		\frac{1}{ \sqrt{\kappa_j} \sqrt{1-\kappa_j} }, 
\label{eq:kappa_prior}
\end{align}
where $a_j =  \tau \sigma^{-1} \sqrt{n} \, s_j$.
When $a_j = 1$, this reduces to $\Beta{\frac{1}{2},\frac{1}{2}}$ which looks like a horseshoe, see Figure~\ref{fig:kappa_prior_hs}.
Thus, a priori, we expect to see both relevant ($\kappa_j=0$, no shrinkage) and irrelevant ($\kappa_j=1$, complete shrinkage) variables.
By changing the value of $\tau$, the prior for $\kappa_j$ places more mass either close to~0~or~1.
For instance, choosing $\tau$ so that $a_j = 0.1$ favors complete shrinkage ($\kappa=1$) and thus we expect more coefficients to be close to zero a priori.
Notice though that for a fixed $\tau$, the sparsity assumptions will be dependent on the input dimension~$D$, and to get around this issue, we need to consider the values of all the shrinkage factors $\kappa_j$ together.
Using this idea, Section~\ref{sec:global_parameter} discusses an intuitive way of designing a prior distribution for $\tau$ based on the assumptions about the number of nonzero components in $\vs \beta$.

Notice also that those variables which vary on larger scale $s_j$ are treated as more relevant a priori, which is the reason why we usually scale all the variables to have unit variance $s_j^2=1$, unless the original scales really carry information about the relevances. 
Another way would be to use the original scales but adjust scales for the local parameters accordingly $\lambda_j \sim \halfCauchy{0, s_j^{-2}}$.

\subsection{Spike-and-slab prior}
\label{sec:spike-and-slab}

The spike-and-slab \citep{mitchell1988,george1993} is a popular shrinkage prior that is often considered as the ``gold standard'' for sparse Bayesian estimation.
The prior is often written as a two-component mixture of Gaussians
\begin{align}
\begin{split}
	\beta_j\given\lambda_j,c,\,\varepsilon &\sim \lambda_j\,\Normal{0,c^2} + (1-\lambda_j)\, \Normal{0,\varepsilon^2}, \\
	\lambda_j &\sim \Ber{\pi}, \qquad j=1,\dots,D,
\end{split}
\label{eq:spike-and-slab}
\end{align}
so that $\varepsilon \ll c$ and the indicator variable $\lambda_j\in \{0,1\}$ denotes whether the coefficient $\beta_j$ is close to zero (comes from the ``spike'', $\lambda_j=0$) or nonzero (comes from the ``slab'', $\lambda_j=1$).
Often we set $\varepsilon = 0$, that is, the spike is taken to be a delta spike at the origin~$\delta_0$, although also $\varepsilon > 0$ could be used~\citep{george1993}.
The user then has to specify the values (or priors) for the slab width $c$ and the prior inclusion probability $\pi$, which encodes the prior information about the sparsity of the coefficient vector $\vs \beta$.
Fixing $c$ is probably the most common approach but by giving it a hyperprior, one can obtain a more heavy-tailed, such as Laplacian, slab \citep{johnstone2004}.

With the choice $\varepsilon=0$, the prior \eqref{eq:spike-and-slab} can be written analogous to \eqref{eq:hs} as
\begin{align}
\begin{split}
	\beta_j \given \lambda_j,c &\sim \Normal{0, c^2 \lambda_j^2}, \\
	\lambda_j &\sim \Ber{\pi}, \qquad j=1,\dots,D,
\end{split}
\label{eq:spike-and-slab2}
\end{align}
so instead of giving continuous priors for $\lambda_j$s as in the horseshoe, here only two values ($\lambda_j = 0,1$) are allowed.
Thus also the shrinkage factor $\kappa_j$ only has mass at $\kappa_j = \frac{1}{1+n \sigma^{-2} s_j^2 c^2}$ and at $\kappa_j = 1$, and the probabilities are $\pi$ and $1-\pi$, respectively.
Letting $c \rightarrow \infty$, all the mass is concentrated at the extremes $\kappa_j=0$ and $\kappa_j=1$, and the resemblance to the horseshoe becomes obvious, see Figure~\ref{fig:kappa_prior_hs}.
Given the similarity of the shrinkage profiles between the horseshoe and spike-and-slab, it is not surprising that the two priors have shown comparable performance in a variety of experiments \citep{carvalho2009,carvalho2010,polson2011}.
The next section discusses an extension of the horseshoe that closely resembles the spike-and-slab prior with a finite slab width $c < \infty$.

\subsection{Regularized horseshoe}
\label{sec:regularized_horseshoe}

As discussed in Section~\ref{sec:horseshoe}, the horseshoe prior favors solutions $\beta_j\approx 0$ and $\beta_j \approx \hat \beta_j$, and it can be shown that under certain conditions, $\bar \beta_j \rightarrow \hat \beta_j$ when $|\hat \beta_j|\rightarrow \infty$ \citep{carvalho2010}.
While this guarantees that the strong signals will not be overshrunk -- and is often considered to be one of the key assets of the prior -- this property can also be harmful, especially when the parameters are weakly identified. 
An example of such case is the flat likelihood arising  in logistic regression with separable data.
As the horseshoe has Cauchy tails, in these problems it suffers basically from the same problems as the Cauchy prior, namely that the posterior means for the regression coefficients may vanish~\citep{ghosh2017}.
Therefore it would be very useful to be able to control the amount of shrinkage for the largest coefficients, which in spike-and-slab prior (Sec.~\ref{sec:spike-and-slab}) is achieved by controlling the slab width.

To guarantee that the prior always shrinks the coefficients at least by a small amount, we introduce the following \emph{regularized horseshoe} prior
\begin{align}
\begin{split}
	\beta_j \given \lambda_j,\tau,c &\sim \Normal{0, \tau^2 \ti \lambda_j^2},  \quad \ti \lambda_j^2 =  \frac{c^2 \lambda_j^2}{c^2 + \tau^2 \lambda_j^2}, \\
	\lambda_j &\sim \halfCauchy{0,1} \,, \quad j=1,\dots,D,
\end{split}
\label{eq:rhs}
\end{align}
where $c > 0$ is a constant that we assume is given for now.
The intuition behind this definition is the following. 
When $\tau^2 \lambda_j^2 \ll c^2$, meaning the coefficient $\beta_j$ is close to zero, then $\ti \lambda_j^2 \rightarrow \lambda_j^2$ and the prior~\eqref{eq:rhs} approaches the original horseshoe.
However, when $\tau^2 \lambda_j^2 \gg c^2$, meaning the coefficient is far from zero, then $\ti \lambda_j^2 \rightarrow c^2/\tau^2$ and the prior~\eqref{eq:rhs} approaches $\Normal{0,c^2}$.
Thus the prior will shrink the small signals as the horseshoe but will also regularize even the largest coefficients as a Gaussian slab with variance $c^2$.

Another way to see this is to notice that the conditional prior for $\beta_j$ can be factored as
\begin{align}
	p(\beta_j \given \lambda_j,\tau,c) 
	\propto \Normal{0, \tau^2 \lambda_j^2} \, \Normal{0,c^2}
	\propto \Normal{0, \tau^2 \ti \lambda_j^2},
\label{eq:rhs2}
\end{align}
from which it is easy to see that depending on the relative magnitudes of $\tau^2 \lambda_j^2$ and $c^2$, the prior operates (roughly) as the narrower one of the two factors.
Therefore the role of $\Normal{0,c^2}$ is to ``soft-truncate'' the extreme tails of the horseshoe, thereby controlling the magnitude of the largest $\beta_j$s.
Letting $c\rightarrow \infty$, we recover the original horseshoe.

\begin{figure}[t]
\centering
	\minput[pdf]{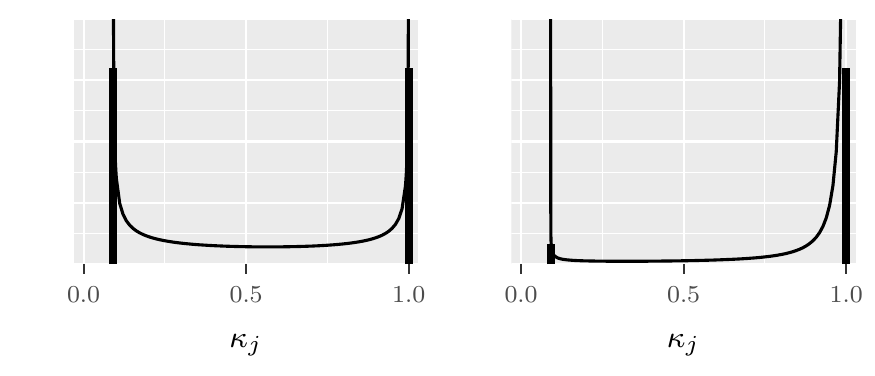}
	\caption{Shrinkage profiles as in Figure~\ref{fig:kappa_prior_hs} but now for the regularized horseshoe~\eqref{eq:rhs} and spike-and-slab~\eqref{eq:spike-and-slab2} with a finite slab width $c=1$. In comparison to Figure~\ref{fig:kappa_prior_hs}, for both priors the first mode is shifted from $\kappa_j=0$ to $\kappa_j=1/(1+ n \sigma^{-2} s_j^2 c^2)$ (for the plots we have selected $n \sigma^{-2} s_j^2 = 10$). }
	\label{fig:kappa_prior_rhs}
\end{figure}

The shrinkage profile of the regularized horseshoe is illustrated in Figure~\ref{fig:kappa_prior_rhs} together with the spike-and-slab with the slab width $c$, which demonstrates the similarity of the two priors.
Using $c < \infty$ has the advantage that it regularizes the parameters $\beta_j$ when they are weakly identified, and allows us also to specify our prior information about the maximum effect $\vc \beta_j$ we expect to see.
The benefit of the proposed approach is illustrated in Section~\ref{sec:separable_data}.

It must be noted that the shrinkage profile in Figure~\ref{fig:kappa_prior_rhs} does not have \emph{exactly} the same shape as the original horseshoe shifted and scaled from interval $(0,1)$ to $(b,0)$ with $b>0$, although it is very close to this.
There is slightly more mass near the left hand side mode, although the difference is too small to be visible in Figure~\ref{fig:kappa_prior_rhs}.
It is possible to retain the exact shape of the horseshoe by defining the modified local parameter $\ti \lambda_j^2$ as
\begin{align}
	\ti \lambda_j^2 =  \frac{ c^2 \lambda_j^2 }{ \frac{\sigma^2}{n s_j^2} + c^2 + \tau^2 \lambda_j^2}.
	\label{eq:lambda_ti}
\end{align}
The details of this result are spelled out in Appendix~\ref{app:rhs}.
The reason why we define the prior as~\eqref{eq:rhs} is that unless $n$ or the slab width $c^2$ is very small, the term $\frac{\sigma^2}{n s_j^2}$ is typically small compared to $c^2$ and leaving it out has little influence in practice.
On the other hand, formulation~\eqref{eq:rhs} is simpler and has the nice interpretation as a product of the original horseshoe and the Gaussian slab~\eqref{eq:rhs2}.
Thus we report the result~\eqref{eq:lambda_ti} only for completeness.

Our formulation requires choosing a value or prior for $c$.
Unless substantial knowledge about the scale of the relevant coefficients exists, we generally recommend placing a prior for $c$ instead of fixing it. 
Often a reasonable choice is 
\begin{align}
	c^2 \sim \InvGamma{\alpha,\beta}, \quad \alpha=\nu/2, \quad \beta=\nu s^2/2,
\label{eq:cprior_ig}
\end{align}
which translates to a Student-$t_\nu(0,s^2)$ slab for the coefficients far from zero and is typically a good default choice for a weakly informative prior.
Another motivation for using inverse-Gamma is that it has a heavy right tail accompanied by a light left tail thereby preventing much mass from accumulating near zero.
This is natural as we do not want to shrink those coefficients heavily towards zero that are already deemed to be far from zero.
Still, we emphasize that out approach is not limited to this choice and also other hyperpriors are possible.
It would also be possible to use variable specific slab widths $c_i$, but we do not explore this further in this paper and leave it for future investigation.

Finally, we would also like to point out that the tail-cutting idea of Equation~\eqref{eq:rhs2} could be used more generally with other priors when relevant.
For instance, we expect our idea  to be useful with the horseshoe+ prior of \cite{bhadra2017} which also has Cauchy tails and therefore suffers from the same problem as the horseshoe.

\subsection{Hierarchical shrinkage}

In addition to the problems with vanishing means, earlier we have also reported sampling issues with the original horseshoe even in simple regression problems~\citep{piironen2015}. 
Technically speaking, the problem arises due to posterior having an extreme funnel shape which is challenging for Markov chain Monte Carlo (MCMC) methods.
The problem was revealed with the help of the divergence diagnostics of the NUTS algorithm \citep{hoffman2014,betancourt2015,betancourt2017a,betancourt2017b} when fitting the models in Stan.

The problem is related to the thick Cauchy tails of the prior, and to overcome the sampling issues, in our technical report we tentatively proposed replacing the half-Cauchy priors for the local parameters~$\lambda_j$ in~\eqref{eq:hs} with half-$t$ priors with small degrees of freedom, such as $\nu=3$, and named this approach the ``hierarchical shrinkage''.
With large enough $\nu$, this seems to help with the sampling issues and remove the divergent transitions produced by NUTS, but the drawback is that the prior becomes less sparsifying.
This is because when the tails of $p(\lambda_j)$ are made slimmer, we need to increase the value for $\tau$ to accommodate large signals, and therefore the prior is not able to shrink the small coefficients efficiently towards zero.
Another limitation of this approach is that in order to fight the problems arising from data separation in logistic regression (Sec.~\ref{sec:regularized_horseshoe}), we would also need to refrain from using half-Cauchy prior for the global parameter~$\tau$ (which we might want to use, see Sec.~\ref{sec:global_parameter}) as this would also lead to Cauchy tails for $p(\beta_j)$.

The less sparsifying nature of the choice $\nu > 1$ will be demonstrated in practice in Section~\ref{sec:separable_data}, where we also show that the regularized horseshoe (Sec.~\ref{sec:regularized_horseshoe}) clearly outperforms this approach.
Thus we no longer recommend increasing the local degrees of freedom, but instead of using the regularized horseshoe.

\section{The global shrinkage parameter}
\label{sec:global_parameter}

This section discusses the prior choice for the global hyperparameter $\tau$.
We begin with a short note on why we prefer full Bayesian inference for $\tau$ over point estimation, and then go on to discuss how we propose to set up the prior $p(\tau)$.

\subsection{Full Bayes versus point estimation}
\label{sec:full_bayes_vs_point_estimation}

In principle, one could use a plug-in estimate for $\tau$, obtained either by cross-validation or maximum marginal likelihood (sometimes referred to as ``empirical Bayes'').
The maximum marginal likelihood estimate has the drawback that it is always in danger of collapsing to $\hat \tau=0$ if the parameter vector happens to be very sparse.
Moreover, rather than being computationally convenient, this approach might actually complicate matters as the marginal likelihood is not analytically available for non-Gaussian likelihoods.
While cross-validation avoids the latter problem and possibly also the first one, it is computationally less efficient than the full Bayesian solution and fails to account for the posterior uncertainty.
For these reasons we recommend full Bayesian inference for $\tau$, and focus on how to specify the prior distribution.

\subsection{Earlier approaches}
\label{sec:earlier_approaches}

\cite{carvalho2009} also recommend full Bayesian inference for $\tau$, and following \cite{gelman2006}, they propose prior
\begin{align}
	\tau \sim \halfCauchy{0,1},
\label{eq:tau_prior_hc1}
\end{align} 
whereas \cite{polson2011} recommend
\begin{align}
	\tau \given \sigma \sim \halfCauchy{0,\sigma^2}.
\label{eq:tau_prior_hcsigma}
\end{align}
If the target variable $y$ is scaled to have marginal variance of one, unless the noise level $\sigma$ is very small, both of these priors typically lead to quite similar posteriors.
However, as we argue in Section~\ref{sec:meff_and_tau}, there is a theoretical justification for letting $\tau$ scale with $\sigma$.
The main motivation for using a half-Cauchy prior for $\tau$ is that it evaluates to a finite positive value at the origin, yielding a proper posterior and allowing even complete shrinkage $\tau \rightarrow 0$, while still having a thick tail which can accommodate a wide range of values.
For these reasons, $\halfCauchy{0,\eta^2}$ is a desirable choice when there are enough observations to let $\tau$ be identified by data. 
Still, we show that in several cases one can clearly benefit by choosing the scale $\eta$ in a more careful manner than simply $\eta=1$ or $\eta=\sigma$, because for most applications these choices place far to much mass for implausibly large values of~$\tau$.
This point is discussed in Section~\ref{sec:meff_and_tau}.
Moreover, the synthetic example in Section~\ref{sec:toy_example} shows that in some cases one could clearly benefit from even more informative prior.  %

\cite{vanDerPas2014} study the optimal selection of $\tau$ in model
\begin{align}
	y_i \sim \beta_i + \varepsilon_i, \quad \varepsilon_i \sim \Normal{0,\sigma^2}, \quad i=1,\dots,n.
\label{eq:model_simple}
\end{align}
They prove that in such a setting, the optimal value (up to a log factor) in terms of mean squared error and posterior contraction rates in comparison to the true~$\vs \beta^*$ is 
\begin{align}
	\tau^* = \frac{p^*}{n},
\label{eq:vanderpas}
\end{align}
where $p^*$ denotes the number of nonzeros in the true coefficient vector $\vs \beta^*$ (assuming such exists).
Their proofs assume that $n,p^* \rightarrow \infty$ and $p^* = o(n)$.
Model~\eqref{eq:model_simple} corresponds to setting $\vc X = \vc I$ and $D=n$ in the usual regression model~\eqref{eq:lgm}.
It is unclear whether and how this result could be extended to a more general $\vc X$, and how one should utilize this result when $p^*$ is unknown (as it usually is in practice).
In section~\ref{sec:meff_and_tau}, we formulate our method of constructing the prior $p(\tau)$ based on the prior information about $p^*$, and show that if $p^*$ was known, our method would also give rise to result~\eqref{eq:vanderpas}, but is more generally applicable.

\subsection{Effective number of nonzero coefficients}
\label{sec:meff_and_tau}

Consider the prior distribution for the shrinkage factor of the $j$th regression coefficient for the linear Gaussian model, Eq.~\eqref{eq:kappa_prior}.
The mean and variance can be shown to be
\begin{align}
	\mean{\kappa_j \given \tau,\sigma} &= \frac{1}{1+a_j}, 
	\label{eq:kappa_mean} \\
	\var{\kappa_j \given \tau,\sigma} &= \frac{a_j }{ 2 (1 + a_j)^2},
	\label{eq:kappa_var}
\end{align}
where $a_j = \tau \sigma^{-1} \sqrt{n} \, s_j$ as earlier.
A given value for the global parameter $\tau$ can be understood intuitively via the prior distribution that it imposes on the effective number of coefficients distinguishable from zero (or effective number of nonzero coefficients, for short), which we define as
\begin{align}
	m_\tx{eff} = \sum_{j=1}^D (1-\kappa_j).
\label{eq:m_eff}
\end{align}
When the shrinkage factors $\kappa_j$ are close to 0 and 1 (as they typically are for the horseshoe prior), this quantity describes essentially how many active or unshrunk variables we have in the model.
It serves therefore as a useful indicator of the effective model size.

Using results~\eqref{eq:kappa_mean} and~\eqref{eq:kappa_var}, the mean and variance of $m_\tx{eff}$ given $\tau$ and $\sigma$ are given by
\begin{align}
	\mean{m_\tx{eff}\given\tau, \sigma} &= \sum_{j=1}^D \frac{a_j}{1+a_j} \\ %
	\var{m_\tx{eff}\given\tau, \sigma} &= \sum_{j=1}^D \frac{a_j }{ 2 (1 + a_j)^2}, %
\end{align}
Let us now assume that, in addition of having a zero mean, each variable also has a unit variance $s_j^2=1$.
In this case the equations above simplify to
\begin{align}
	\mean{m_\tx{eff}\given\tau, \sigma} &= \frac{\tau \sigma^{-1}\sqrt{n} }{1+\tau \sigma^{-1} \sqrt{n}} D, \label{eq:meff_mean} \\
	\var{m_\tx{eff}\given\tau, \sigma} &= \frac{\tau \sigma^{-1} \sqrt{n} }{ 2 (1 + \tau \sigma^{-1} \sqrt{n})^2} D. \label{eq:meff_var}
\end{align}
The expression for the mean~\eqref{eq:meff_mean} is helpful.
First, from this expression it is evident that to keep our prior information about $m_\tx{eff}$ consistent, $\tau$ must scale as $\sigma/\sqrt{n}$.
Priors that fail to do so, such as~\eqref{eq:tau_prior_hc1}, favor models of varying size depending on the noise level $\sigma$ and the number of data points $ n$. 
Second, if our prior guess for the number of relevant variables is $p_0$, it is reasonable to choose the prior so that most of the prior mass is located near the value
\begin{align}
	\tau_0 =  \frac{p_0}{D-p_0} \frac{\sigma}{\sqrt{n}},
\label{eq:tau0}
\end{align}
which is obtained by solving equation $\mean{m_\tx{eff}\given \tau,\sigma} = p_0$.
Note that this is typically quite far from $1$ or $\sigma$, which are used as scales for priors~\eqref{eq:tau_prior_hc1} and~\eqref{eq:tau_prior_hcsigma}.
For instance, if $D = 1000$ and $n = 200$, then prior guess $p_0=5$ gives about $\tau_0=3.6\cdot10^{-4} \sigma$.

To further develop the intuition about the connection between $\tau$ and $m_\tx{eff}$, it is helpful to visualize the prior imposed on $m_\tx{eff}$ for different prior choices for $\tau$.
This is most conveniently done by drawing samples for $m_\tx{eff}$\,; we first draw $\tau \sim p(\tau)$ and $\lambda_1,\dots,\lambda_D \sim \halfCauchy{0,1}$, then compute the shrinkage factors $\kappa_1,\dots,\kappa_D$ from ~\eqref{eq:kappa}, and finally $m_\tx{eff}$ from~\eqref{eq:m_eff}. 

\begin{figure*}[t]
\centering
	\setlength{\figureheight}{0.3\textwidth}
	\setlength{\figurewidth}{0.85\textwidth}
	\pgfplotsset{
	compat=newest,
	major tick length={0.03cm},
	x tick label style={font=\tiny},
	y tick label style={font=\tiny},
	ylabel style={text width=5em, rotate=-90, align=left},
	legend style={font=\tiny},
	} 
	\minput[pdf]{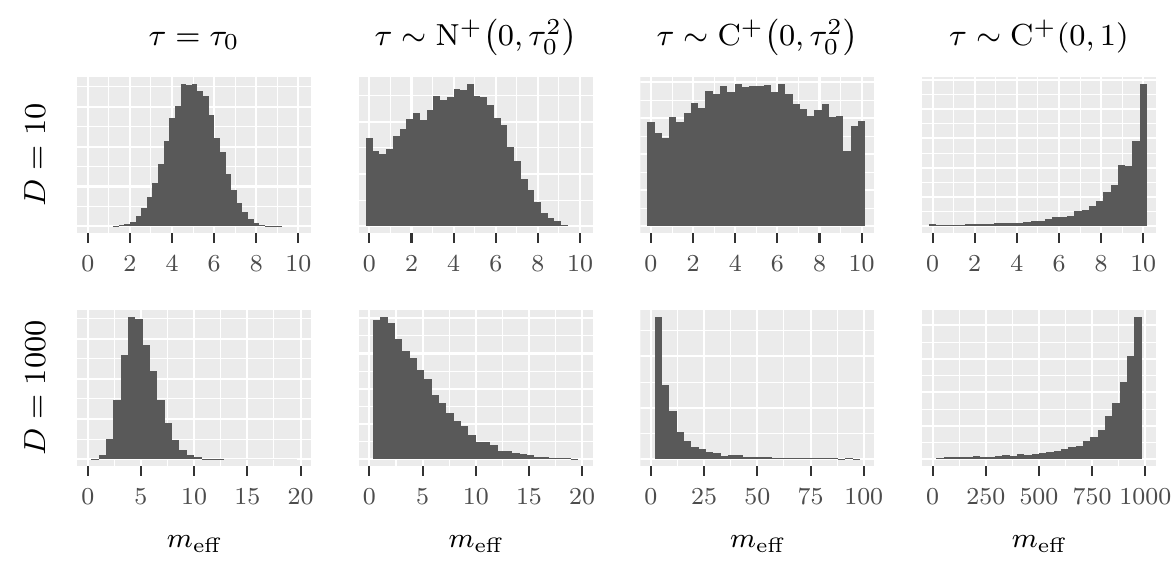}
	\caption{Histograms of prior draws for $m_\tx{eff}$ (effective number of nonzero regression coefficients, Eq.~\eqref{eq:m_eff}) imposed by different prior choices for $\tau$, when the total number of input variables is $D=10$ and $D=1000$. $\tau_0$ is computed from formula~\eqref{eq:tau0} assuming $n=100$ observations with $\sigma=1$ and $p_0 = 5$ as the prior guess for the number of relevant variables. Note the varying scales on the horizontal axes in the bottom row plots.}
	\label{fig:meff_prior}
\end{figure*}

Figure~\ref{fig:meff_prior} shows histograms of prior draws for $m_\tx{eff}$ for some different prior choices for $\tau$, with total number of variables $D=10$ and $D=1000$, assuming $n=100$ observations with $\sigma = 1$.
The first three priors utilize the value $\tau_0$ which is computed from~\eqref{eq:tau0} using $p_0 = 5$ as our hypothetical prior guess for the number of relevant variables.
Fixing $\tau=\tau_0$ results in a nearly symmetric distribution around $p_0$, while a half-normal prior with scale $\tau_0$ yields a skewed distribution favoring solutions with $m_\tx{eff} < p_0$ but allowing larger values to also be accommodated.
The half-Cauchy prior behaves similarly to the half-normal, but results in a distribution with a much thicker tail giving substantial mass also to values much larger than $p_0$ when $D$ is large. 
Figure~\ref{fig:meff_prior} also illustrates why prior $\tau \sim \halfCauchy{0,1}$ is often a dubious choice: it places far too much mass on large values of $\tau$, consequently favoring solutions with most of the coefficients unshrunk.
Thus when only a small number of the variables are relevant -- as we typically assume -- this prior results in sensible inference only when $\tau$ is strongly identified by data.
Notice also that, if we changed the value of $\sigma$ or $n$, the first three priors for $\tau$ would still impose the same prior for $m_\tx{eff}$, but this is not true for $\tau\sim \halfCauchy{0,1}$.

This way, by studying the prior for $m_\tx{eff}$, one can easily choose the prior for  $\tau$ based on the information about the number of nonzero parameters.
Because the prior information can vary substantially for different problems and the results depend on the information carried by the data, there is no globally optimal prior for $\tau$ that works for every single problem.
Some recommendations, however, will be given in Section~\ref{sec:recommendations} based on these theoretical considerations and experiments presented in Section~\ref{sec:experiments}.

We conclude by pointing out a connection between our reference value~\eqref{eq:tau0} and the oracle result~\eqref{eq:vanderpas} for the simplified model~\eqref{eq:model_simple}. 
As pointed out in the last section, model~\eqref{eq:model_simple} corresponds to setting $\vc X = \vc I$ (which implies $n=D$ and $\vc X^\tp \vc X = \vc I$) in the usual regression model~\eqref{eq:lgm}.
Using this fact and repeating the steps needed to arrive at~\eqref{eq:tau0}, we get 
\begin{align}
	\tau_0 = \frac{p_0}{D-p_0} \sigma.
\label{eq:tau0_simple}
\end{align}
Suppose now that we select $p_0 = p^*$, that is, our prior guess is oracle.
Using the same assumptions as \cite{vanDerPas2014}, namely that $n,p^* \rightarrow \infty$ and $p^* = o(n)$, and additionally that $\sigma = 1$, we get $\tau_0 \rightarrow p^* / D = \tau^*$. This result is natural, as it means it is optimal to choose $\tau$ so that the imposed prior for the effective number of nonzero coefficients $m_\tx{eff}$ is centered at the true number of nonzeros $p^*$. 
This further motivates why $m_\tx{eff}$ is a useful quantity.

\subsection{Regularized horseshoe and other shrinkage priors}

As discussed in Section~\ref{sec:regularized_horseshoe}, when we set $c < \infty$, the shrinkage profile of the regularized horseshoe~\eqref{eq:rhs} becomes approximately equivalent to that of the horseshoe shifted from interval $(0,1)$ to $(b_j,1)$, where $b_j = \frac{1}{1+n \sigma^{-2} s_j^2 c^2}$.
Thus the shrinkage factor under the regularized horseshoe satisfies approximately $\ti \kappa_j = (1-b_j) \kappa_j + b_j$, where $\kappa_j$ denotes the shrinkage factor for the original horseshoe.
From this we get $1-\ti\kappa_j = (1-b_j)(1-\kappa_j)$.
Assuming further that all the variables have a unit variance $s_j^2 = 1$ and thus $b_j=b=\frac{1}{1+n\sigma^{-2} c^2}$, the effective model complexity under the regularized horseshoe satisfies
\begin{align*}
	\ti m_\tx{eff} =  (1-b) m_\tx{eff},
\end{align*}
where $m_\tx{eff}$ is the effective number of nonzeros for the original horseshoe.
Thus with a given $\tau$, the effective complexity for the regularized horseshoe is always less than for the pure horseshoe, because those coefficients that are far from zero are still affected by the slab.
Therefore we can naturally use result~\eqref{eq:tau0} also for the regularized horseshoe with $p_0$ as our prior guess for the number of coefficients far from zero, but remembering that those coefficients will experience the regularization by the slab.

The concept of effective number of nonzeros could also be used with shrinkage priors other than the (regularized) horseshoe, as long as the prior can be written as a scale mixture of Gaussians like~\eqref{eq:hs}. 
Many such alternatives have been proposed, including the double-exponential or Laplace~\citep{park2008}, Dirichlet-Laplace~\citep{bhattacharya2015},  R-square induced Dirichlet decomposition \citep{zhang2016}, and horseshoe+~\citep{bhadra2017}.
For instance, the Dirichlet-Laplace prior has a Dirichlet concentration hyperparameter $a$ that strongly affects the sparsity properties of the prior, and based on the experiments of \cite{bhattacharya2015} has a substantial effect on the results.
It would therefore be interesting the investigate a prior information calibrated selection of $a$ using our framework.

Depending on the prior, corresponding analytical results like~\eqref{eq:meff_mean} and~\eqref{eq:meff_var} may or may not be available, but as long as one is able to sample both from $p(\tau)$ and $p(\lambda_j)$, it is always easy to draw samples from the prior distribution for $m_\tx{eff}$, and therefore investigate the effect of hyperprior or hyperparameter choice on the effective model complexity.
It must be noted though, that for those prior for which the shrinkage factors $\kappa$ are not near 0 or 1, the values of $m_\tx{eff}$ can be more difficult to interpret.

\subsection{Non-Gaussian observation models}
\label{sec:nongaussian_lik}

When the observation model is non-Gaussian, the exact analysis from Section~\ref{sec:meff_and_tau} is analytically intractable.
We can, however, perform the analysis using a Gaussian approximation to the likelihood.
Using the second order Taylor expansion for the log likelihood, the approximate posterior for the regression coefficients given the hyperparameters becomes
\begin{align*}
	p(\vs \beta \given \vs \Lambda, \tau,\phi,\mc D) &\approx \Normal{\vs \beta \given \vs {\bar \beta}, \vs \Sigma}, \\
	\vs {\bar \beta} &= \tau^2 \vs \Lambda \left( \tau^2 \vs \Lambda + (\vc X^\tp \vs {\ti \Sigma}^{-1} \vc X)^{-1} \right)^{-1} \vs {\hat \beta}, \\
	\vs \Sigma &= (\tau^{-2}\vs \Lambda^{-1} + \vc X^\tp \vs {\ti \Sigma}^{-1} \vc X)^{-1},
\end{align*}
where $\vc {\ti z} = (\ti z_1,\dots,\ti z_n)$, $\vs {\ti \Sigma} = \tx{diag}(\ti \sigma_1^2,\dots,\ti \sigma_n^2)$ and $\vs {\hat \beta} = (\vc X^\tp \vs {\ti \Sigma}^{-1} \vc X)^{-1} \vc X^\tp \vs {\ti \Sigma}^{-1} \vc {\ti z}$ (assuming the first inverse  exists).
Here $\phi$ denotes the possible dispersion parameter and $(\ti z_i,\ti \sigma_i^2)$ the location and variance for the $i$th Gaussian pseudo-observation.
These are obtained from the first and second order derivatives of the log-likelihood terms $L_i(f_i,\phi)$ with respect to the linear predictor $f_i = \vs \beta^\tp \vc x_i$ at the posterior mode $\bar f_i = \vs {\bar \beta}^\tp \vc x_i$ \citep[ch.~16.2]{gelman2013book}
\begin{align*}
	\ti z_i &= \bar f_i - \frac{L_i'(\bar f_i,\phi)} {L_i''(\bar f_i,\phi)}, \quad
	\ti \sigma^2_i = - \frac{1} {L_i''(\bar f_i,\phi)}.
\end{align*}
The fact that some of the observations are more informative than others -- meaning $\ti \sigma_i^2$ is not constant -- makes further simplification somewhat difficult.

To proceed, we make the rough assumption that we can replace each $\ti \sigma_i^2$ by a single variance term $\ti \sigma^2$.
Assuming further that the covariates are uncorrelated with zero mean and variances $\var{x_j}=s_j^2$ (as in Sec.~\ref{sec:meff_and_tau}), the posterior mean for the $j$th coefficient satisfies $\bar \beta_j = (1-\kappa_j)\hat \beta_j$ with shrinkage factor given by
\begin{align}
	\kappa_j = \frac{1}{1 + n \ti \sigma^{-2} \tau^2 s_j^2 \lambda_j^2}.
\label{eq:kappa_nongaussian}
\end{align}
The discussion in Section~\ref{sec:meff_and_tau} therefore also approximately holds for the non-Gaussian observation model, except that $\sigma^2$ is replaced by $\ti \sigma^2$.
Still, this leaves us with the question, which value to choose for $\ti \sigma^2$ to exploit this result in practice?

\begin{table}[t]%
\abovetopsep=2pt
\caption{The pseudo variances for the most commonly used generalized linear models to be used as approximate plug-in values for $\sigma^2$ in equations of Section~\ref{sec:meff_and_tau}. In practice when necessary, we usually replace $\mu$ by the sample mean $\bar y$. In the Gamma distribution, $\alpha$ denotes the shape parameter so that $\var{y}=\mu^2/\alpha$. In the inverse Gaussian, $\lambda$ is the shape parameter so that $\var{y}=\mu^3/\lambda$. See \cite{mccullagh1989book}.}
\label{tab:pseudovar}
\centering
\begin{tabular}{ lllccc }
\toprule
Model & Link & $\doo \mu / \doo f$ & $V(\mu)$ & $A(\phi)$ & Pseudo variance $\ti \sigma^2$   \\ 
\midrule
Gaussian & Identity & $1$ & $1$ & $\sigma^2$ & $\sigma^2$  \\ 
Binomial & Logit & $\mu(1-\mu)$ & $\mu(1-\mu)$ & $1$ & $\mu^{-1}(1-\mu)^{-1}$  \\ 
Poisson 	& Log & $\mu$ & $\mu$ & $1$ & $\mu^{-1}$ \\
Gamma 		& Inverse & $-\mu^2$ & $\mu^2$ & $\alpha^{-1}$ & $\mu^{-2} \alpha^{-1}$ \\
Inverse Gaussian & Inverse squared & $-\mu^3/2$ & $\mu^3$ & $\lambda^{-1}$ & $4 \mu^{-3} \lambda^{-1}$ \\ 
\bottomrule
\end{tabular}
\end{table}

For the generalized linear models with $y$ having a distribution in the exponential family with natural parameter $\theta$ and dispersion $\phi$, the log likelihood for a single observation has the form \citep{mccullagh1989book}
\begin{align*}
	L = \frac{y\theta - B(\theta)}{A(\phi)} - C(y,\phi),
\end{align*}
for some specific functions $A(\cdot)$, $B(\cdot)$ and $C(\cdot)$.
The pseudo variance for a given observation $y$ is then (see Appendix~\ref{app:pseudovariance})
\begin{align}
	\ti \sigma^2 
	= - \frac{1}{L''} 
	=  A(\phi) \left[ 
	\frac{1}{V(\mu)} \left( \frac{\doo \mu}{\doo f}\right)^2 
	- (y-\mu) \frac{\doo}{\doo f} \left( \frac{1}{V(\mu)} 
	  \frac{\doo \mu}{\doo f} \right)
	\right]^{-1},
\label{eq:pseudovar}
\end{align}
where $V(\mu) = B''(\theta)$ is the variance function, $\mu = \mean{y} = B'(\theta)$ the expected value. %
The simplified expressions for the most commonly used generalized linear models with their canonical links are listed in Table~\ref{tab:pseudovar}.

We observe that $\ti \sigma^2$ is a product between $A(\phi)$ and a term that in general depends on  $f$, $\mu$, and $y$, although for canonical links the dependence from $y$ vanishes because the derivative of  $(\frac{\doo\mu}{\doo f}) / V(\mu)$ with respect to $f$ is zero \citep[ch.~2]{mccullagh1989book}.
Thus it makes sense that for those non-Gaussian models that have a dispersion parameter $\phi$ (like Gamma and inverse Gaussian models), the pseudo variance and therefore also $\tau$ should scale with $A(\phi)$.
For the non-constant multiplier of $A(\phi)$ in~\eqref{eq:pseudovar} we can use, for example, value obtained by setting $\mu$ equal to the sample mean of $\bar y$.
This approach, although crude, seems to work reasonably well in practice.
For instance, in binary classification, if we have the same number of observations from both classes, then $\mu=0.5$, yielding $\ti \sigma^2=4$ which was observed to give good results in our earlier study~\citep{piironen2017b}.

\subsection{More complex models}
\label{sec:other_models}

Although we limit our discussion to generalized linear models, different authors have employed horseshoe prior in various other models.
For instance, \cite{faulkner2017} consider trend filtering for modelling time series and use horseshoe as a sparsifying prior on the $k$th-order forward differences to express prior assumptions about the number of rapid changes in the underlying signal.
As another example, \cite{ghosh2017b} use horseshoe prior over the weights in  Bayesian neural networks to effectively turn off some of the nodes in the network.

When the model gets more complicated, one cannot simply use the reference result~\eqref{eq:tau0} to guide the hyperprior choice because it is derived assuming the linear model~\eqref{eq:lgm}.
Unless similar results can easily be derived for the model of interest,  we recommend a pragmatic approach of drawing from the prior for different values of $\tau$ and studying the effect on the sparsity (how many coefficients fall below a certain threshold) to get an idea of a reasonable range of values.
Although this strategy may seem crude, we argue that it should still be better than not thinking about the prior at all.
Moreover, our results  (Sec.~\ref{sec:real_world_data}) suggest that with a weakly informative prior such as ${\tau \sim \halfCauchy{0,\tau_0^2}}$, having even a rough ballpark figure of the correct magnitude for $\tau_0$ can already be a clear improvement compared to the simple $\tau \sim \halfCauchy{0,1}$.

\section{Experiments}
\label{sec:experiments}

This section illustrates the benefit of the theoretical advances on synthetic and real world data.
All considered models were fitted using Stan\footnote{The experiments with the original horseshoe on the real world data  are taken from \cite{piironen2017b} and were run using Stan version~2.12.0, whereas the experiments with the regularized horseshoe are run using newer version~2.15.1} (codes in the supplementary material) with the default settings unless otherwise stated, running 4 chains, 2000 samples each, first halves discarded as warmup.

\subsection{Synthetic data}

\subsubsection{Toy example}
\label{sec:toy_example}

\begin{figure*}[t]
	\centering
	\minput[pdf]{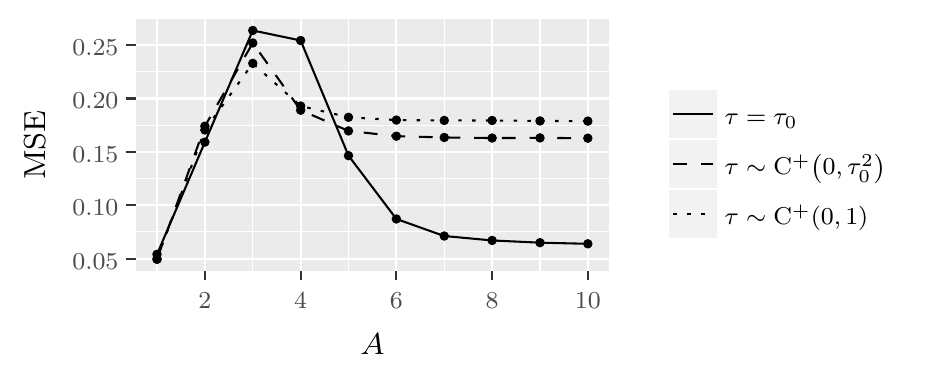}
	\caption{\emph{Toy example}: Mean squared error (MSE) between the estimated and the true coefficient vector of length $n=400$ on average over 100 different data realizations. The true coefficient vector has $p_*=20$ elements with a nonzero value equal to $A$ and the rest are zeros.}
	\label{fig:toyexample_mse}
\end{figure*}
\begin{figure*}[t]
	\centering
	\minput[pdf]{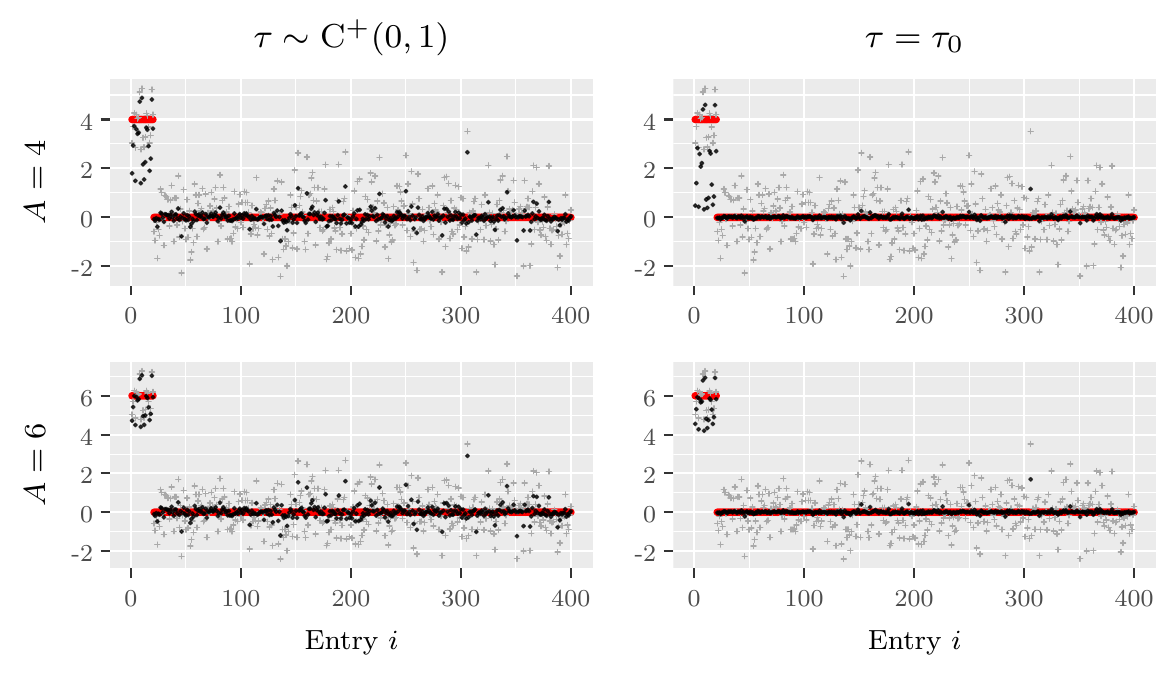}
	\caption{\emph{Toy example}: An example data realization $\vc y=(y_1,\dots,y_n)$ (gray crosses), posterior mean~$\vs{\bar \beta}=(\bar \beta_1,\dots, \bar \beta_n)$ (black dots) and the true signal $\vs \beta_*$ (red lines) for $A=4$ (top row) and $A=6$ (bottom row). In both cases the oracle value for $\tau$ helps to shrink the zero components in $\vs \beta$ but also overshrinks the actual signals in the case $A=4$.}
	\label{fig:toyexample_coeff}
\end{figure*}
\begin{figure*}[t]
	\centering
	\minput[pdf]{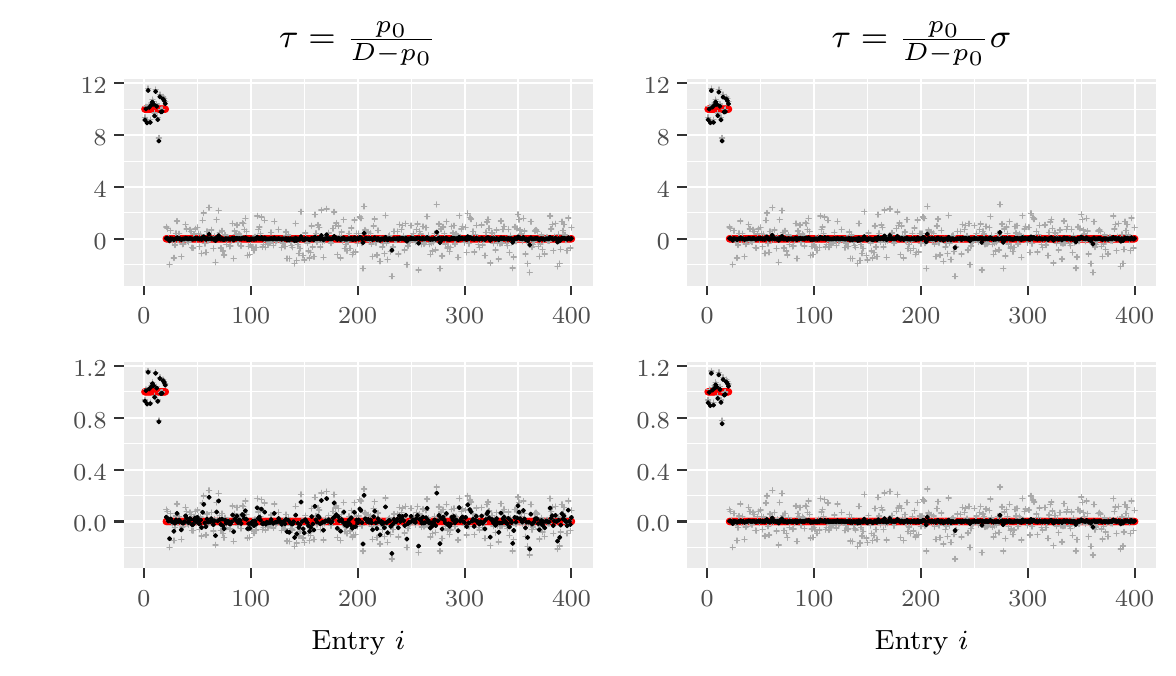}
	\caption{\emph{Toy example}: Top row: one data realization $\vc y$ (gray crosses), the posterior mean estimates $\vs {\bar \beta}$ (black dots) and the true signal $\vs \beta_*$ (red lines) when $A=10$ and the global parameter is fixed either to $\tau = \frac{p_0}{D-p_0}$ or $\tau = \frac{p_0}{D-p_0}\sigma$. Bottom row: Otherwise the same but now the scale of the observations is changed by multiplying them by $0.1$.}
	\label{fig:toyexample_scaling}
\end{figure*}

We first illustrate the impact of the hyperprior choice $p(\tau)$ with a toy example similar to the one discussed by \cite{vanDerPas2014}.
Consider model~\eqref{eq:model_simple}, where each $y_i$ is generated by adding Gaussian noise with $\sigma^2=1$ to the corresponding signal~$\beta_i$.
We generated 100 data realizations with $n=400$ and the true $\vs \beta_*$ having $p_*=20$ nonzero entries equal to $A=1,2,\dots,10$ with the rest of the entries being zeros.
We then computed the mean squared error (MSE) between the estimated posterior mean $\vs{ \bar \beta}$ and the true $\vs \beta_*$ assuming the pure horseshoe prior with hyperpriors $\tau\sim \halfCauchy{0,1}$, $\tau\sim \halfCauchy{0,\tau_0^2}$ and $\tau=\tau_0$, where $\tau_0$ is calculated from Equation~\eqref{eq:tau0_simple} with the oracle prior guess $p_0=p_*$.
Although the last choice reflects stronger prior information than we would typically expect to have in practice, the purpose of this setup is simply to demonstrate that one can substantially improve the inferences using our framework provided substantial prior knowledge exists.
Notice though, that even when we set $\tau=\tau_0$, we do not treat $\tau$ as completely fixed, because it depends on~$\sigma$ which is treated as an unknown parameter with vague prior $p(\sigma^2) \propto \sigma^{-2}$.

Figure~\ref{fig:toyexample_mse} shows the MSE for the three hyperpriors for different values of $A$.
For each prior, the error is largest close to $A = \sqrt{2\log 400} \approx 3.5$, which is called the ``universal threshold" for this problem \citep{johnstone2004, vanDerPas2014}.
Below this threshold the nonzero components in $\vs \beta$ are too small to be detected and are thus shrunk too heavily towards zero which introduces error.
For $A=4$ the informative $\tau=\tau_0$ actually yields the worst results due to this overshrinkage (see discussion below), but gives clearly superior results for larger $A$.
The choice $\tau\sim \halfCauchy{0,\tau_0^2}$ gives better results than $\tau\sim \halfCauchy{0,1}$ but is clearly inferior to  $\tau=\tau_0$.

Figure~\ref{fig:toyexample_coeff} illustrates the data $\vc y$ and the estimated coefficients $\vs{\bar \beta}$ for one particular data realization when $A=4$ and $A=6$.
In both cases the informative choice $\tau = \tau_0$ helps to shrink the zero components in $\vs \beta$ towards zero, but for $A=4$ also overshrinks the nonzero components. 
The reason for the overshrinkage is that some observations $y_i$ that correspond to zero signal $\beta_i=0$ happen to have similar magnitude to the observations coming from a nonzero signal $\beta_i=A$, and thus these irrelevant components ``steal" from the limited budget for $m_\tx{eff}$.
For this particular value of $A$ the overshrinkage of the actual signals happens to be worse in terms of MSE than undershrinkage of the zero components, and thus one would get better results by setting $p_0$ to be slightly above the true $p_*$ (results not shown).
For $A=6$ the actual signals are large enough to be distinguished from zero, and the informative selection of $\tau$ yields substantially better estimate for $\vs \beta$.

Finally, Figure~\ref{fig:toyexample_scaling} illustrates the importance of scaling $\tau$ with the noise level~$\sigma$.
The top row shows one data realization and the posterior mean estimates $\vs {\bar \beta}$ when the global parameter is fixed either to $\tau = \frac{p_0}{D-p_0}$ or to $\tau = \frac{p_0}{D-p_0}\sigma$.
For these data both yield essentially the same result, since the true noise variance $\sigma^2$ is one.
However, when the observations are scaled by multiplying them by $0.1$ (bottom row), the value for $\tau$ that does not scale with $\sigma$ yields clearly worse results than in the first case, while the results for the latter value remain practically unchanged.
What essentially happens is that when the observations are transformed to a smaller scale, then fixing $\tau = \frac{p_0}{n-p_0}$ increases the prior expectation for $m_\tx{eff}$ by the same factor (10 in this case) and thus it will favor solutions with many more coefficients far from zero.

This small experiment has relevance also regarding the more general model~\eqref{eq:lgm} because in the ideal case of uncorrelated predictors with unit variance ${\vc X^\tp \vc X = n \vc I}$, we can think that we have a single observation of each $\beta_j$ with variance $\sigma^2/n$.
In practice these conditions are rarely met but the idea is still useful.\sloppy

\subsubsection{Classification with separable data}
\label{sec:separable_data}

%
{
\begin{figure}[t]
\hspace{-2em}
	\minput[pdf]{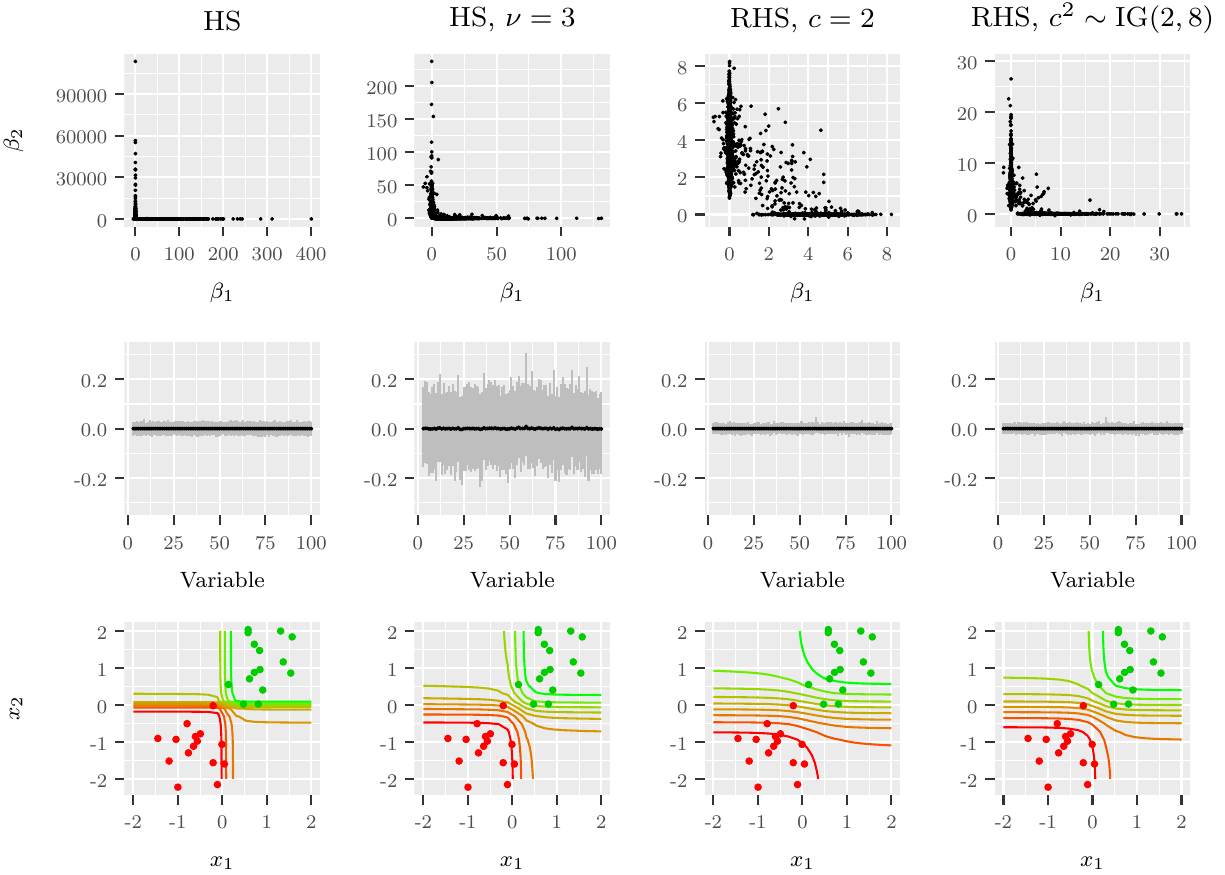}
	\caption{\emph{Separable classification}. Top row: draws for the regression coefficients of the two relevant features. Middle row: median and 80\%-interval for the coefficients of the irrelevant features. Bottom row: the observed data (red and green denoting the different classes) and the predictive class probabilities for $\ti y = 1$ as a function of the first two inputs, given that the rest of the inputs are set to zero. First two columns denote the results for the pure horseshoe and hierarchical shrinkage, and the last two columns for the regularized horseshoe. }
	\label{fig:separable}
\end{figure}}

The purpose of this example is to illustrate the problem with the original horseshoe not controlling the magnitude of the largest coefficients and show how the regularized horseshoe can solve this issue.
We generated $n=30$ binary classification observations so that for the instances in the first class, the first two features were drawn from a Gaussian with mean $1$ and scale $0.5$, whereas in the other class the mean of the first two features was $-1$ (the data are visualized in Fig.~\ref{fig:separable}).
In addition, we generated 98 irrelevant features drawn from the standard Gaussian, so that the total input dimension was $D=100$.
We then fitted the standard logistic regression model with prior $\beta_0 \sim \Normal{0,5^2}$ for the intercept and four different priors for the regression coefficients $\beta_j$: the original horseshoe, hierarchical shrinkage with $\nu=3$, and regularized horseshoe with slab scales $c=2$ and $c^2 \sim \InvGamma{2,8}$, the latter of which corresponds to a Student-$\Student[4]{0,2^2}$ slab (see Sec.~\ref{sec:regularized_horseshoe}).
In each case we used hyperprior $\tau \sim \halfStudent[3]{0,\tau_0^2}$ (the conclusions of this example are not sensitive to this choice). For consistent sparsity assumptions, for the original and the regularized horseshoe, $\tau_0$ was calculated from~\eqref{eq:tau0} using $p_0=2$, and for the hierarchical shrinkage by solving numerically $\mean{m_\tx{eff}\given \tau=\tau_0} = p_0 = 2$.

Figure~\ref{fig:separable} shows the scatter plots of the posterior draws for $\beta_1$ and $\beta_2$ (top row) as well as the medians and 80\%-intervals for $\beta_3,\dots,\beta_{100}$ (middle row), for the four different priors.
Because the data are separable using only $x_2$, this feature is a ``solitary separator'' and thus the mean for $\beta_2$ does not exist under the horseshoe prior due to its Cauchy tails~\citep{ghosh2017}.
Although for $\beta_1$ the mean exist ($x_1$ is not a solitary separator), the posterior has substantial mass for very large values for this parameter also.
Moreover, for the original horseshoe the NUTS produces almost 200 divergent transitions after the warmup showing clear problems with sampling.
Using $\nu=3$ for the local parameters cuts down the tails and reduces the number of divergent transitions to a few, but still yields quite fat posterior tails for these two coefficients and results in much less shrinkage for the coefficients of the irrelevant features (middle row).
Finally, the regularized horseshoe exhibits the most satisfactory performance cutting down the tails for $\beta_1$ and $\beta_2$ while still being able to shrink the irrelevant coefficients as well as the horseshoe.
In this case, fixing the scale of the Gaussian slab to $c=2$ results in too strong regularization for the relevant coefficients and generally we would prefer a less informative choice such as $c \sim \InvGamma{2,8}$, but our purpose here was to demonstrate how easy it is to specify different levels of regularization for largest coefficients using the new prior~\eqref{eq:rhs}.

The top row of Figure~\ref{fig:separable} clearly reveals the multimodality of the posterior for $\beta_1$ and $\beta_2$.
This does not produce marked problems in this case (e.g. MCMC convergence problems) but in general the multimodality can be an issue both for the original and the regularized horseshoe, and we will discuss this further in Sections~\ref{sec:real_world_data}~and~\ref{sec:discussion}.

\subsection{Real world data -- microarray cancer classification}
\label{sec:real_world_data}

\begin{table}[t]%
\centering
\abovetopsep=2pt
\caption{Summary of the real world microarray cancer datasets; number of predictor variables $D$ and dataset size $n$.} %
\label{tab:datasets}
\begin{tabular}{ llccc }
\toprule
Dataset & Type & $D$ & $n$ \\ 
\midrule
Ovarian & Binary classif. & 1536 & 54   \\
Colon & Binary classif. & 2000 & 62   \\
Prostate & Binary classif. & 5966 & 102   \\
Leukemia (ALL-AML) & Binary classif. & 7129 & 72   \\
\bottomrule
\end{tabular}
\end{table}

This section further illustrates the important concepts with some real world examples.
We use the four microarray cancer classification datasets from our earlier paper~\citep{piironen2017b}.
The datasets are summarized in Table~\ref{tab:datasets} and can be found online.\footnote{Colon, Prostate and Leukemia: \url{http://featureselection.asu.edu/datasets.php}; Ovarian data: request from the first author if needed.}
For all the datasets we used the standard logistic regression model with a vague prior $\beta_0 \sim \Normal{0,10^2}$ for the intercept, and the original and regularized horseshoe priors for the regression coefficients to compare the differences between the two.
For these problems, we reduced the number of draws per chain to 1000 to reduce the computation time.
\begin{figure}[t]
\centering
	\minput[pdf]{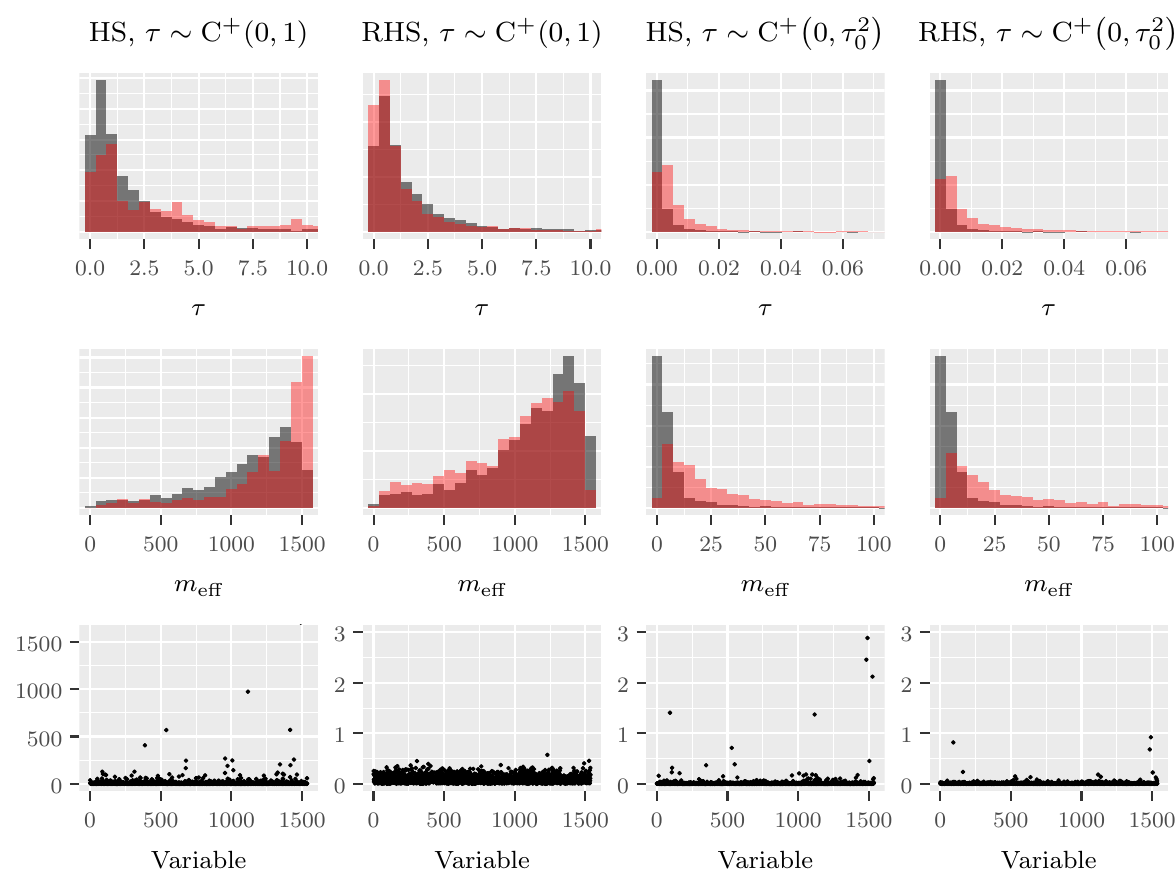}
	\caption{\emph{Ovarian dataset}\,: Histograms of prior (grey) and posterior (red) draws for $\tau$ (top row) and $m_\tx{eff}$ (middle row), and absolute values of the posterior means for the regression coefficients $|\bar\beta_j|$ (bottom row) imposed by different prior choices. The first two columns denote the results for the pure and regularized horseshoe with hyperprior $\tau \sim \halfCauchy{0,1}$, and the last two columns the same but with $\tau \sim \halfCauchy{0,\tau_0^2}$, where $\tau_0$ is computed from~\eqref{eq:tau0} with prior guess $p_0=3$.}
	\label{fig:ovarian}
\end{figure}
\begin{figure}[t]
\centering
	\minput[pdf]{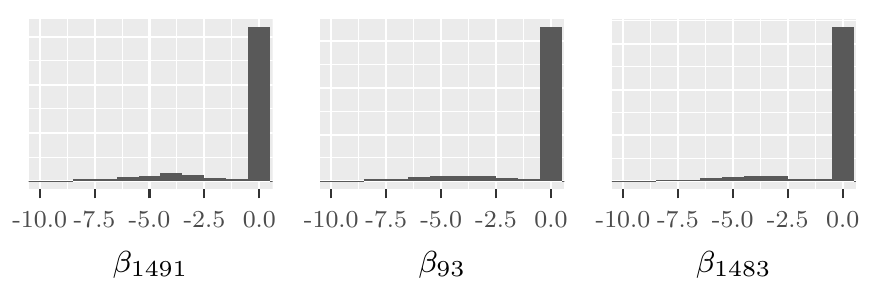}
	\caption{\emph{Ovarian dataset}\,: Histograms of the posterior draws under the regularized horseshoe prior for the three regression coefficients with the largest absolute mean. The histograms highlight the multimodality of the posterior.}
	\label{fig:ovarian_coeff}
\end{figure}
\begin{figure*}[t]
\centering
	\minput[pdf]{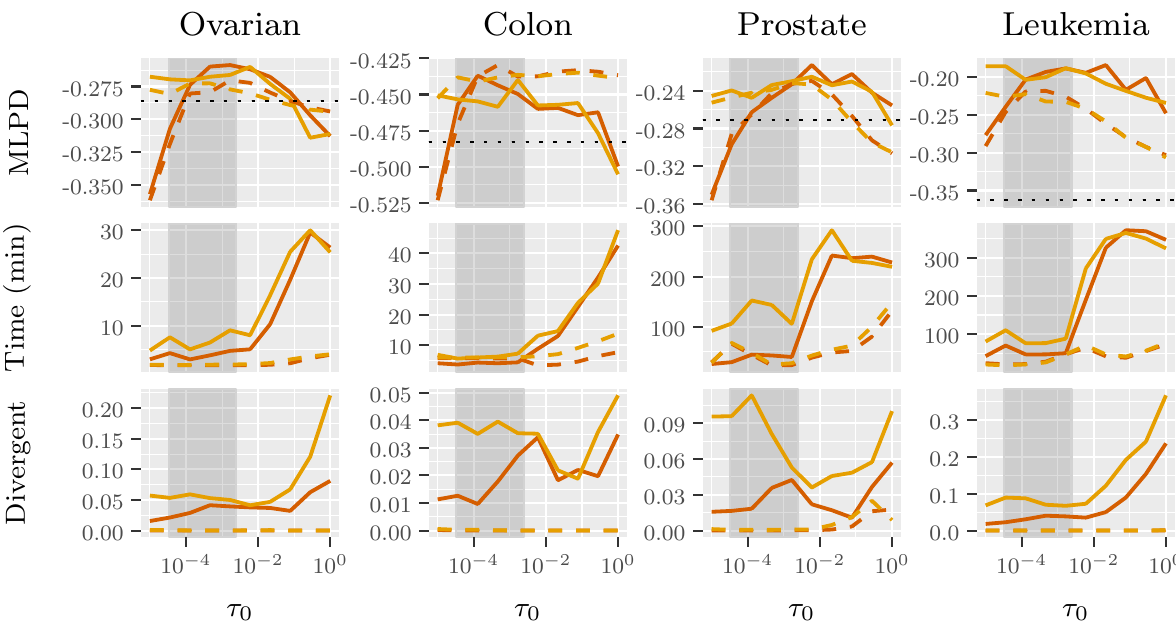}
	\caption{\emph{Microarray datasets}\,: Mean log predictive density (MLPD) on test data (top row), computation time (middle row), and the fraction of divergent transitions after warmup (bottom row) as a function of $\tau_0$ for two hyperpriors:  $\tau \sim \halfNormal{0,\tau_0^2}$ (red), and $\tau \sim \halfCauchy{0,\tau_0^2}$ (yellow). Solid lines denote the pure horseshoe and dashed lines the regularized horseshoe with $c^2\sim \InvGamma{2,8}$. The shaded area denotes the values for $\tau_0$ that correspond to sparsity guesses between $p_0=1$ and $p_0 = n$ (Eq.~\eqref{eq:tau0}). The black dotted line in the top row plots shows the MLPD for Lasso. All the results are averaged over 50 random splits into training and test sets.}
	\label{fig:results_classif}
\end{figure*}
We first consider the Ovarian dataset as a representative example of how the prior choice $p(\tau)$ and the use of the regularized horseshoe can affect the results. 
We fitted the model to the data with two hyperprior choices, $\tau \sim \halfCauchy{0,1}$ and $\tau \sim \halfCauchy{0,\tau_0^2}$, where $\tau_0$ is computed from~\eqref{eq:tau0} using $p_0 = 3$ as our prior guess for the number of relevant variables.
For the regularized horseshoe we used hyperprior $c^2 \sim \InvGamma{2,8}$ (as with the synthetic example, Sec.~\ref{sec:separable_data}) which corresponds to a Student-$\Student[4]{0,2^2}$ slab (see Sec.~\ref{sec:regularized_horseshoe}).

Figure~\ref{fig:ovarian} shows prior and posterior draws for $\tau$ and $m_\tx{eff}$, and the absolute values of the posterior means for the regression coefficients, for the different prior configurations.
The results for $\tau \sim \halfCauchy{0,1}$ illustrate how weakly $\tau$ is identified by the data: there is very little difference between the prior and posterior samples for $\tau$ and consequently for $m_\tx{eff}$, and thus this ``non-informative'' prior actually has a strong influence on the posterior.
With the pure horseshoe this results in severe under-regularization and implausibly large magnitude for the regression coefficients.
This happens because the effective model complexity is so large that the classes become separable, and due to the vanishing maximum likelihood estimate also the posterior mean vanishes (as in the synthetic example in Sec.~\ref{sec:separable_data}).
With the regularized horseshoe, even with the complete separation, the magnitude of the coefficients remains sensible due to the regularization by the slab. 

Replacing the scale of the half-Cauchy hyperprior with $\tau_0$, reflecting a more sensible guess for the number of relevant variables, has a substantial effect on the posterior: the posterior mass for $m_\tx{eff}$ becomes concentrated on much smaller values and the magnitude of the regression coefficients more sensible.
For this hyperprior choice the difference between the original and regularized horseshoe is much less severe, but also in this case the largest coefficients are smaller for the regularized horseshoe.
How this affects the predictive accuracy will be discussed in a moment.

A potential explanation for why $\tau$ and therefore $m_\tx{eff}$ are not strongly identified here is that there are a lot of correlations in the data.
For instance, the predictor $j=1491$ which appears relevant based on its regression coefficient, has an absolute correlation of at least $|\rho|=0.5$ with 314 other variables, out of which 65 correlations exceed $|\rho|=0.7$.
This indicates that there are a lot of potentially relevant but redundant predictors in the data, and thus similar fit could be obtained by models with varying levels of sparsity.

Another difficulty with correlating predictors is that they cause the posterior of the regression coefficients to become multimodal, which usually results in convergence issues.
Figure~\ref{fig:ovarian_coeff} shows the posterior draws under the regularized horseshoe prior for the three coefficients with the largest absolute mean.
All three coefficients have most of the draws near zero but some of the draws are far from zero, illustrating the multimodality.
The $\hat R$-values \citep[potential scale reduction factor, see e.g.,][ch.~11]{gelman2013book} for these parameters were $1.09$, $1.20$ and $1.05$ indicating problems with the convergence of the MCMC chains.
Although these convergence issues do not seem to have a drastic negative impact on the predictive accuracy (see the discussion below),  the multimodality dramatically reduces the sampling efficiency and is therefore a major practical and theoretical concern and deserves more attention in the future research (see also the discussion in Section~\ref{sec:discussion}).

To investigate the effect of the prior choices on the prediction accuracy, we split each dataset into two halves, using one fifth of the data as a test set.
All the results were then averaged over 50 such random splits into training and test sets.
We carried out the tests for priors $\tau \sim \halfCauchy{0,\tau_0^2}$ and $\tau \sim \halfNormal{0,\tau_0^2}$ with various $\tau_0$.
The half-normal prior was included in the tests to investigate whether it sometimes could be beneficial to have a strong control over $\tau$ by using a short-tailed prior.
To get a baseline for the comparisons, we also computed the prediction accuracies to Lasso with the regularization parameter tuned by 10-fold cross-validation.
The Lasso results were computed with the default settings of the R-package {\tt glmnet}~\citep{friedman2010}. %

The top row of Figure~\ref{fig:results_classif} shows the effect of the prior choice on the test prediction accuracy, and the other two rows the computation time\footnote{Wall time when the four chains are run in parallel using four cores.} and the fraction of divergent transitions produced by the NUTS after the warm-up.
For the original horseshoe (solid lines) the results illustrate a clear benefit from using even a crude prior guess for the number of relevant variables $p_0$: transforming any guess between $p_0=1$ and $p_0 = n$ into a value of $\tau_0$ using Equation~\eqref{eq:tau0} (shaded region) and using this as the scale for the half-Cauchy prior instead of $\tau_0 = 1$ yields improved prediction accuracy and reduced computation time in all the datasets. %
The regularized horseshoe seems in general less sensitive to the prior choice for $\tau$ (Ovarian and Colon datasets), but in some cases clearly benefits from a more carefully chosen prior (Prostate and Leukemia datasets).
For both priors, using half-Cauchy hyperprior for $\tau$ is clearly less sensitive to the prior guess $\tau_0$, and yields better results than the half-normal especially when $\tau_0$ was chosen to be too small.

In terms of the predictive accuracy, the regularized horseshoe performs overall comparably to the original horseshoe: for the Colon dataset it performs slightly better, for Leukemia slightly worse, and for Ovarian and Prostate about the same as the pure horseshoe.
Only for the Colon and Leukemia datasets the difference is statistically significant (the errorbars are left out from the plot to avoid mess).
In Leukemia dataset where the regularized horseshoe loses to the pure horseshoe the chosen prior for the slab width $c$ is probably unnecessarily restrictive leading to too large regularization for the largest coefficients.
It is evident that in this case using a looser prior for $c$ could improve the results (because when $c\rightarrow \infty$ we recover the original horseshoe), but we feel that these results are enough to convince the reader that a reasonably chosen hyperprior $p(c)$ does not necessarily compromise and in some cases can even improve the accuracy compared to the pure horseshoe.

On the other hand, the regularized horseshoe clearly improves the sampling robustness of the posterior.
The regularized horseshoe produces only very few divergent transitions after the warm-up if any (the fraction is non-negligible only for the Prostate dataset with a poorly chosen global prior $p(\tau)$), whereas for the original horseshoe the fraction varies between $1$--$30\%$ which is really a lot.
With this many divergences, there is always a concern about a biased inference and this cannot be taken too lightly (see the next Section for recommendations).
Moreover, also the computation times are systematically either smaller or similar to the original horseshoe, which is due to better behaving posterior.

For any reasonably selected prior (shaded region), the (regularized) horseshoe consistently outperforms Lasso in terms of predictive accuracy, but in some cases the difference is not very large. 
A clear advantage for Lasso, on the other hand, is that it is hugely faster, with computation time of less than a second for these problems, whereas even for the regularized horseshoe, the computation starts to get quite involved for the two biggest problems (around 30--60 minutes for Prostate and Leukemia datasets), which is the price for the full Bayesian inference.

\section{Recommendations}
\label{sec:recommendations}

Based on the theoretical considerations and the experimental results, instead of the original horseshoe, we recommend using the regularized horseshoe with a weakly informative prior on $c$, such as \eqref{eq:cprior_ig} with appropriately chosen scale $s$ and degrees of freedom $\nu$.
As illustrated in Section~\ref{sec:experiments}, this does not compromise the predictive accuracy but greatly improves the sampling robustness of the prior and is resistant to problems originating from weak identifiability of the parameters.
Still, it is worthwhile to compare the results to a model with a relatively loose prior on $c$  using standard model assessment techniques \citep{vehtari2012,vehtari2017} to get an indication if the slab width was chosen to be too restrictive.
However, even if a very large value for $c$ yields better predictive fit, if it produces a large number of divergent transitions, it cannot be recommended because in such cases the inference is likely to have a bias with an unknown magnitude~\citep{betancourt2017a}.
In our experience, for the regularized horseshoe it is often possible to get rid off the divergences by tuning Stan's adaptation routine as explained by \cite{betancourt2017b}.

Also, instead of using the simple $\tau \sim \halfCauchy{0,1}$, for linear regression we generally recommend a  weakly informative default choice $\tau \given \sigma \sim \halfCauchy{0,\tau_0^2}$, where $\tau_0$ is computed from~\eqref{eq:tau0} using the prior guess $p_0$ for the number of relevant variables.
For the other generalized linear models (GLMs), an approximately equivalent choice is obtained by replacing $\sigma$ with a appropriate plug-in value as explained in Section.~\ref{sec:nongaussian_lik}.
Based on the results on the real world data, this choice seems to perform well unless $p_0$ is chosen to be much too large.
However, the toy example shows that sometimes even better results can be obtained by more informative prior.

\section{Discussion}
\label{sec:discussion}

This paper has discussed the use of the horseshoe prior for sparse Bayesian generalized linear models.
We considered two methodological advances.
First, we proposed a generalization of the horseshoe -- called the regularized horseshoe -- that operates otherwise similarly as the horseshoe but allows specifying the regularization to the coefficients that are far from zero.
Second, we introduced a new concept -- effective number of nonzero parameters -- which is useful for guiding the hyperprior choice for the global shrinkage parameter.

The experiments demonstrated the benefit of both of these approaches.
The ability to regularize those parameters that are far from zero is useful especially when the parameters are only weakly identified by the data.
As an example we discussed the logistic regression with data separation. 
Adding a small regularization to the largest coefficients ensures that the posterior mean will exist, leading to more reasonable parameter estimates and faster posterior exploration.
The regularized horseshoe solves also the problems with the divergent transitions that have previously been an indication about problems in posterior simulation and possibly biased inference~\citep{betancourt2017b}.

Regarding the hyperprior choice for the global shrinkage parameter, we argued that the previous default choices are often dubious based on their tendency to favor solutions with too many parameters unshrunk.
The experiments show that for many datasets, one can obtain clear improvements -- in terms of better parameter estimates, prediction accuracy and faster computation -- by coming up even with a crude guess for the number of relevant variables and transforming this knowledge into a prior for $\tau$ using our proposed framework.
Based on our results, for a reasonably selected global hyperprior, the (regularized) horseshoe outperforms Lasso in terms of predictive accuracy.
A notable difference is that Lasso produces a truly sparse solution with exact zeros for some coefficients, whereas horseshoe does not.
After fitting the full model, a truly sparse solution without losing predictive accuracy could be obtained using the projective variable selection~\citep{piironen2017a}\footnote{Code available at~\url{https://github.com/stan-dev/projpred}.}.

Although these advances improve the overall performance and applicability of the horseshoe prior to various problems, some challenges still persist.
When using the horseshoe prior with correlating predictors, the major concern is always the multimodality of the posterior, which can lead to difficulties in sampling and especially to slow convergence of the MCMC.
The experiments indicated that the regularized horseshoe can improve the sampling robustness  but does not remove the multimodality.
It must be noted though, that the multimodality is a direct consequence of the sparsifying prior assumption that favors solutions where only one in a group of correlating predictors would have a nonzero coefficient.
Whether this assumption is reasonable in practice is a more fundamental question.
It could make sense -- both computationally and theoretically -- to employ a horseshoe or other sparsifying prior on some transformed set of features -- for instance principal components of the original predictors -- instead of the original variables themselves.
This would yield a unimodal posterior but the effects on the other aspects such as the predictive accuracy remain to be explored.
We leave these ideas for future investigation.

\appendix

\section{Derivation of the regularized horseshoe}
\label{app:rhs}

The regularized horseshoe from Section~\ref{sec:regularized_horseshoe} can be derived by designing a new prior with shrinkage profile otherwise similar to the horseshoe, but so that instead of favoring $\kappa_j=0$ and $\kappa_j=1$, the prior favors $\kappa_j=b_j$ and $\kappa_j=1$.
This is achieved by defining the shrinkage factor for the new prior as
\begin{align}
	\ti \kappa_j = (1-b_j) \kappa_j + b_j,
\end{align}
where $\kappa_j$ is the shrinkage factor for the original horseshoe.
Now $\ti \kappa_j \rightarrow 1$ when $\kappa_j \rightarrow 1$, but $\ti \kappa_j \rightarrow b_j$ when $\kappa_j \rightarrow 0$, so we shift the shrinkage profile of the horseshoe from interval $(0,1)$ to $(b_j,0)$.
By plugging in the expression for the shrinkage factor $\kappa_j = 1/(1+a_j^2 \lambda_j^2)$ where $a_j^2 = n \sigma^{-2} \tau^2 \, s_j^2$, with a straightforward manipulation we get
\begin{align}
	\ti \kappa_j = \frac{1 + b_j \, a_j^2 \lambda_j^2}{1 + a_j^2\lambda_j^2}.
\end{align}
We want to write this in the form $\ti \kappa_j = 1/(1+a_j^2 \ti \lambda_j^2)$.
Solving $\ti \lambda_j^2$ from this equation yields
\begin{align}
	\ti \lambda_j^2 = \frac{(1-b_j)\lambda_j^2}{1 + b_j \,a_j^2\lambda_j^2}.
\label{aeq:lambda_tilde_raw}
\end{align}
For convenience, we require that $b_j=\frac{1}{1+n \sigma^{-2} s_j^2 c^2}$ which corresponds to shrinkage by Gaussian with variance $c^2$. 
By plugging this into~\eqref{aeq:lambda_tilde_raw}, after a few lines of straightforward algebra we are left with 
\begin{align}
	\ti \lambda_j^2 
	= \frac{c^2 \lambda_j^2}{ \frac{\sigma^2}{n s_j^2} + c^2 + \tau^2\lambda_j^2}.
\label{aeq:lambda_tilde}
\end{align}
Thus we have shown that the prior
\begin{align}
	\beta_j \given \lambda_j,\tau \sim \Normal{0, \tau^2 \ti \lambda_j^2}, 
	\quad \lambda_j \sim \halfCauchy{0,1},
\label{aeq:rhs}
\end{align}
with $\ti \lambda_j^2$ defined by~\eqref{aeq:lambda_tilde} has the shrinkage profile of the horseshoe shifted from the interval $(0,1)$ to $(b_j,1)$, where $b_j=\frac{1}{1+n \sigma^{-2} s_j^2 c^2}$.
As discussed in Section~\ref{sec:regularized_horseshoe}, the term $\frac{\sigma^2}{n s_j^2}$ is typically small compared to $c^2$ so by leaving this out from~\eqref{aeq:rhs}, we get the prior~\eqref{eq:rhs}.

\section{Pseudo variance for Non-Gaussian observations}
\label{app:pseudovariance}

As discussed in Section~\ref{sec:nongaussian_lik}, assuming $y$ has a distribution in the exponential family with natural parameter $\theta$ and dispersion $\phi$, the log likelihood has the form
\begin{align*}
	L = \frac{y\theta - B(\theta)}{A(\phi)} - C(y_i,\phi),
\end{align*}
for some specific functions $A(\cdot)$, $B(\cdot)$ and $C(\cdot)$.
From the well-known relations 
\begin{align*}
	\mean{\frac{\doo L}{\doo \theta}} = 0, \quad \tx{and} \quad
	\mean{\frac{\doo^2 L}{\doo \theta^2} } + \mean{ \left( \frac{\doo L}{\doo \theta} \right)^2 } = 0,
\end{align*}
we can easily derive
\begin{align*}
	\mean{y} &=  B'(\theta), \\
	\var{y} &= B''(\theta) A(\phi).
\end{align*}
We denote $\mu = B'(\theta)$ and $V(\mu) = B''(\theta)$, from which we can also deduce $\frac{\doo \mu}{\doo \theta} = V(\mu)$.
The first derivative is given by the chain rule
\begin{align*}
	\frac{\doo L}{\doo f} 
	= \frac{\doo L}{\doo \theta}\, \frac{\doo \theta}{\doo \mu}\, \frac{\doo \mu}{\doo f} 
	= \frac{y-\mu}{A(\phi)}\, \frac{1}{V(\mu)}\, \frac{\doo\mu }{ \doo f}.
\end{align*}
The second derivative is then
\begin{align*}
	\frac{\doo^2 L}{\doo f^2} 
	= \frac{1}{A(\phi)} \left[ (y-\mu) \frac{\doo}{\doo f} \left( \frac{1}{V(\mu)} 
	  \frac{\doo \mu}{\doo f} \right) 
	 - \frac{1}{V(\mu)} 
	 \left( \frac{\doo \mu}{\doo f}\right)^2 \right].
\end{align*}
Thus the pseudo variance becomes
\begin{align*}
	\ti \sigma^2 
	= - \left( \frac{\doo^2 L}{\doo f^2} \right)^{-1}
	=  A(\phi) \left[ 
	\frac{1}{V(\mu)} \left( \frac{\doo \mu}{\doo f}\right)^2 
	- (y-\mu) \frac{\doo}{\doo f} \left( \frac{1}{V(\mu)} 
	  \frac{\doo \mu}{\doo f} \right)
	\right]^{-1}.
\end{align*}

\section{Stan codes}

\subsection{Simple parametrization}
\label{app:rhs_code_param1}

The following shows the Stan code for the linear Gaussian model with the regularized horseshoe prior using a straightforward parametrization.
In our experience this code works fine, but in Appendix~\ref{app:rhs_code_param2} we also provide another code using different parametrization (with which we generated our results).
This is worth trying if the simple code has issues with sampling (produces divergences).

In the code, both $\tau$ and $\lambda_j$ are given half-$t$ priors with the degrees of freedom and the scale defined by the user (the scale can be adjusted only for $\tau$).
Setting {\tt nu\_local} $=1$ corresponds to the horseshoe.
{\tt nu\_global} $ = 1$ gives $\tau$ a half-Cauchy prior.
The scale for $\tau$ is {\tt scale\_global*sigma}, so if we want to set this to be $\tau_0=\frac{p_0}{D-p_0} \frac{\sigma}{\sqrt{n}}$ (Eq.~\eqref{eq:tau0}), we should set \texttt{scale\_global} $ = \frac{p_0}{(D-p_0)\sqrt{n}}$. 
The code assumes a Student-$t$ slab for regularizing the largest coefficients, and the scale and degrees of freedom can be specified using {\tt slab\_scale} and {\tt slab\_df} arguments.

\vspace{1em}
\lstset{language=C, tabsize=2, basicstyle=\ttfamily\scriptsize, literate={~} {$\sim$}{1}} 
\lstinputlisting{\codesdir glm_gauss_rhs_literal.stan}

The code for the logistic regression model is very similar, we simply remove the lines related to the noise deviation {\tt sigma}, and change the observation model and the type of the target variable data {\tt y}.
Notice also that now the scale for $\tau$ is simply {\tt scale\_global}.
Thus, to follow our recommendation, we set {\tt scale\_global} $=\tau_0=\frac{p_0}{D-p_0} \frac{\sigma}{\sqrt{n}}$ (Eq.~\eqref{eq:tau0}), by using a suitable plug-in value for $\sigma$ (Sec.~\ref{sec:nongaussian_lik}).
The lines that need to be changed (in addition to removing definitions related to {\tt sigma}) are shown below.

\lstset{language=C, tabsize=2, basicstyle=\ttfamily\scriptsize, literate={~} {$\sim$}{1}} 
\lstinputlisting{\codesdir glm_ber_rhs_diff.stan}

\subsection{Alternative parametrization}
\label{app:rhs_code_param2}

This parameterization was proposed by \cite{peltola2014} (codes at \url{https://github.com/to-mi/stan-survival-shrinkage}).
In practice we have not observed problems with the code presented in Appendix~\ref{app:rhs_code_param1} but if it has issues with sampling, it is worth trying the following code (using which we ran our experiments).
Below is the code for the Gaussian observation model.

\lstset{language=C, tabsize=2, basicstyle=\ttfamily\scriptsize, literate={~} {$\sim$}{1}} 
\lstinputlisting{\codesdir glm_gauss_rhs.stan}

Again, the code for the logistic regression model is very similar.
Below are the lines that need to be changed (in addition to removing the definitions related to {\tt sigma}.

\lstset{language=C, tabsize=2, basicstyle=\ttfamily\scriptsize, literate={~} {$\sim$}{1}}
\lstinputlisting{\codesdir glm_ber_rhs_diff.stan}

\section{{\tt rstanarm} code}

The horseshoe prior is implemented in the R-package {\tt rstanarm}, which contains a lot of precompiled Stan code and provides an easy-to-use interface to the most commonly used regression models.
At the time of writing this, the package does not implement the regularized horseshoe but we hope it will in the near future.

Assuming the predictor matrix and the corresponding targets are loaded into variables {\tt x} and {\tt y}, the linear Gaussian model with the horseshoe prior can be fitted as follows:
\lstset{language=C, tabsize=2, basicstyle=\ttfamily\scriptsize, literate={~} {$\sim$}{1}}
\lstinputlisting{\codesdir rstanarm_lm.R}
The other generalized linear models can be fitted in a similar manner.
Here is an example of fitting a logistic regression model (assume now that values {\tt y} are binary):
\lstset{language=C, tabsize=2, basicstyle=\ttfamily\scriptsize, literate={~} {$\sim$}{1}}
\lstinputlisting{\codesdir rstanarm_glm.R}

\section*{Acknowledgments}

We thank Andrew Gelman, Michael Betancourt and Daniel Simpson for helpful comments for improving the manuscript.
We also acknowledge the computational resources provided by the Aalto Science-IT project.

\bibliographystyle{imsart-nameyear}
\bibliography{references}

\end{document}